\documentclass[twocolumn,showpacs,preprintnumbers,amsmath,amssymb,pra,unsortedaddress,superscriptaddress,longbibliography,letterpaper,floatfix]{revtex4-2}
\usepackage{xcolor}
\usepackage{graphicx}
\usepackage{dcolumn}
\usepackage{bm}
\usepackage{braket}
\usepackage{verbatim} 
\usepackage[english]{babel}
\usepackage{xfrac}
\usepackage{soul}

\begin{document}


\title{Quantum Many-Body Scars in Few-Body Dipole-Dipole Interactions}

\author{Sarah E. Spielman}
\affiliation{Department of Physics, Bryn Mawr College, Bryn Mawr, PA 19010.}
\author{Alicia Handian}%
\affiliation{Department of Physics and Astronomy, Ursinus College, Collegeville, PA 19426.}

\author{Nina P. Inman}%
\affiliation{Department of Physics, Bryn Mawr College, Bryn Mawr, PA 19010.}
\affiliation{Department of Physics and Astronomy, Rice University, Houston, Texas, 77251.}

\author{Thomas J. Carroll}
\affiliation{Department of Physics and Astronomy, Ursinus College, Collegeville, PA 19426.}

\author{Michael W. Noel}%
\affiliation{Department of Physics, Bryn Mawr College, Bryn Mawr, PA 19010.}

\date{\today}

\begin{abstract}
We simulate the dynamics of Rydberg atoms resonantly exchanging energy via \mbox{two-}, \mbox{three-}, and four-body dipole-dipole interactions in a one-dimensional array. Using simplified models of a realistic experimental system, we study the initial state survival probability, mean level spacing, spread of entanglement, and properties of the energy eigenstates. By exploring a range of disorders and interaction strengths, we find regions in parameter space where the \mbox{three-} and four-body dynamics either fail to thermalize or do so slowly. The interplay between the stronger hopping and weaker field-tuned interactions gives rise to quantum many-body scar states, which play a critical role in slowing the dynamics of the \mbox{three-} and four-body interactions.
\end{abstract}


\maketitle

\section{Introduction}

The thermodynamic evolution of strongly interacting, closed quantum systems offers insight into behavior that cannot be described by conventional statistical mechanics. Unlike classical systems, which explore all of their states with equal probability in the thermodynamic limit, the evolution of an isolated quantum system is unitary. The eigenstate thermalization hypothesis (ETH) proposes that if a quantum system thermalizes for any initial state, then it must likewise reach thermal equilibrium when initialized in any of the many-body energy eigenstates~\cite{deutsch_quantum_1991,srednicki_chaos_1994,rigol_thermalization_2008}. For a system that obeys the ETH, entanglement can spread rapidly while the memory of the initial state becomes inaccessible to local measurements. In the presence of disorder, an alternative fate is many-body localization (MBL), in which some signature of the initial state remains at long times~\cite{nandkishore_manybody_2015,abanin_recent_2017, abanin_colloquium_2019}. Nonergodic dynamics are also possible if the initial state has significant overlap with only a small subset of the energy eigenstates. These special eigenstates were recently dubbed \textit{quantum many-body scar states}~\cite{turner_weak_2018}. 


Understanding these dynamics will provide insight into the fundamental physics of energy transport, localization, collective behavior, and entanglement growth. Over the past two decades, significant numerical and theoretical progress has been made~\cite{oganesyan_localization_2007,pal_manybody_2010,serbyn_local_2013,huse_phenomenology_2014,znidaric_manybody_2008,bardarson_unbounded_2012,kim_testing_2014,celardo_shielding_2016,serbyn_criterion_2015,santos_cooperative_2016,tavora_inevitable_2016,tavora_powerlaw_2017,wang_manybody_2021,bhakuni_stability_2020,geraedts_characterizing_2017,nosov_correlationinduced_2019,sala_ergodicity_2020,yao_manybody_2014,nandkishore_manybody_2017,singh_effect_2017,nag_manybody_2019,prasad_manybody_2021,thomson_quasimanybody_2020}. Furthermore, the development of experimental platforms using superconducting qubits~\cite{guo_observation_2021,gong_experimental_2021,chen_observation_2021,zhu_observation_2022}, trapped ions~\cite{richerme_nonlocal_2014,smith_manybody_2016}, and nitrogen-vacancy centers~\cite{kucsko_critical_2018} has made it possible to explore these dynamics in the laboratory.


Control of closed quantum systems is necessary for engineering quantum devices and for the analog modeling of condensed matter. The ability to precisely control their geometry and interactions makes ultracold Rydberg atoms an excellent candidate for studying the quantum dynamics of closed systems. Ultracold Rydberg atoms have been used to explore thermalization, the spread of entanglement, and quantum many-body scar states~\cite{orioli_relaxation_2018,whitlock_diffusive_2019,sous_possible_2018,sous_manybody_2019,signoles_glassy_2021,lukin_probing_2019,lippe_experimental_2021,bernien_probing_2017a,turner_weak_2018}.

Resonant F\"{o}rster, or dipole-dipole, interactions among ultracold Rydberg atoms have been studied extensively~\cite{anderson_resonant_1998,lukin_dipole_2001,afrousheh_determination_2006,reinhard_effect_2008,van_ditzhuijzen_spatially_2008,younge_mesoscopic_2009,altiere_dipoledipole_2011,kutteruf_probing_2012,gunter_observing_2013,ravets_coherent_2014,pillet_rydberg_2016a}. In a two-body resonant interaction, the initial and final atomic pair states are degenerate and coupled by the dipole-dipole operator $\hat{V}_{dd}\propto 1/R^3$, where $R$ is the distance between the two atoms. In this work, we consider two types of resonant dipole-dipole interactions.

\begin{figure}
    \centering
    \includegraphics{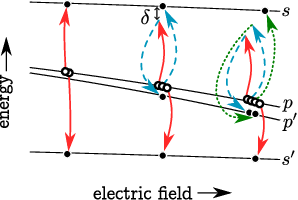}
    \caption{\mbox{Two-}, \mbox{three-}, and four-body interactions for the four-level $spp's'$ model. The horizontal axis is the applied electric field while the vertical axis is energy. Pairs of solid red, dashed blue, and dotted green arrows represent individual detuned steps in the \mbox{three-} and four-body interactions. Open circles indicate initial states and closed circles indicate final states. While these circles are horizontally offset to allow space for the transition arrows, individual resonant interactions occur at only one electric field. The detuning $\delta$ of the three-body interaction is indicated by a double-arrow at the top-center. These energy exchanges are described by Eqs.~(\ref{eq:spps-two})-(\ref{eq:spps-ar}) and the Hamiltonians are given by Eq.~(\ref{eq:ham3-spps}), Eq.~(\ref{eq:ham2-spps}), and Eq.~(\ref{eq:ham4-spps}).}
    \label{fig:spps-model}
\end{figure}

If each final electronic state is different from each initial electronic state, then the interaction is typically tuned into resonance by using an electric field to adjust the Stark energy levels. This defines a ``field-tuned'' interaction in which the interaction is resonant at a specific field. In rubidium, for example, the $36p_{3/2,|m_j|=1/2} + 36p_{3/2,|m_j|=1/2} \leftrightarrow 36s_{1/2} + 37s_{1/2}$ exchange is resonant at an electric field of 3.29~V/cm~\cite{liu_time_2020}, as shown by the leftmost set of arrows in Fig.~\ref{fig:spps-model}.

Resonant field-tuned interactions can also arise when there are degenerate few-body initial and final electronic states, if they are coupled by detuned intermediate states. \mbox{Three-} and four-body dipole-dipole interactions have been observed and studied in both cesium and rubidium Rydberg atoms~\cite{gurian_observation_2012,faoro_borromean_2015,tretyakov_observation_2017,liu_time_2020,beterov_fast_2018,ryabtsev_coherence_2018,cheinet_threebody_2020}. Though recent work has shown that localization can persist in the presence of three-body interactions~\cite{yusipov_quantum_2017,mujal_fewboson_2019}, most studies of thermalization to date focus on two-body interactions.






There is also a set of interactions that do not depend on the electric field tuning, in which the electronic states are swapped between initial and final states. For example, the $39p_{3/2} + 39s_{1/2} \leftrightarrow 39s_{1/2} + 39p_{3/2}$ exchange is always resonant in a spatially homogeneous field~\cite{fahey_imaging_2015}. These interactions are often referred to as hopping interactions, since one can imagine the upper energy state to be an excitation that hops from atom to atom. 


Recently, Liu \textit{et al.}, including some of the authors of this work, measured the time evolution of \mbox{two-}, \mbox{three-}, and four-body interactions in rubidium for a frozen Rydberg gas~\cite{liu_time_2020}. The experiment was a typical quantum quench in which the atoms were excited to the same initial Rydberg state. An electric field then tuned dipole-dipole interactions into resonance, which coupled the atoms to nearby final Rydberg states. The measured fraction of atoms in the upper final Rydberg state appeared to grow more slowly than expected for the \mbox{three-} and four-body interactions.

Motivated by these results, we have constructed an idealized model of few-body dipole-dipole interactions. We numerically solve the Schr\"{o}dinger equation using exact diagonalization for one-dimensional chains of Rydberg atoms. Since our simulations are performed at the resonant electric fields, we assume that energy levels not involved in the interactions can be neglected. In order to study the dynamics at longer times and to maximize the size of the system, we ignore atomic motion~\cite{li_dipoledipole_2005} and the finite lifetime of the Rydberg atoms~\cite{branden_radiative_2009,marcassa_measurement_2009}. We also ignore the angular dependence~\cite{carroll_angular_2004,ravets_measurement_2015}. Each of these are expected to play a role in a real experimental system.

In our simulations, we vary both the interaction energy and the spatial disorder. When only two-body interactions are present, we find that the system thermalizes rapidly in all cases. However, for \mbox{three-} and four-body interactions (alongside the two-body hopping interactions), the system evolves slowly for a large region of the parameter space of interaction energy and disorder. We identify two reasons for this. First, we find numerical evidence for quantum many-body scar states. Second, increasing the disorder slows the dynamics and suggests that these systems could be candidates for MBL studies.

The quantum many-body scars arise when the two-body hopping interactions are significantly stronger than the \mbox{three-} or four-body field-tuned interactions, which is true for a wide range of interatomic spacings. This is similar to the context in which quantum many-body scars were first discovered. In the Rydberg system studied by Turner \textit{et al.}, the coupling between the ground state and Rydberg state was much weaker than the van der Waals interaction between Rydberg states~\cite{bernien_probing_2017a,turner_weak_2018}. 

We consider two models of the dipole-dipole energy exchange. First, we use a model that includes the four relevant energy levels in the experiment of  Liu \textit{et al.}, labeled $s$, $p$, $p'$, and $s'$, as shown in Fig.~\ref{fig:spps-model}~\cite{liu_time_2020}. We shall refer to this as the $spp's'$ model. In Sec.~\ref{sec:spps}, we show that quantum many-body scar states emerge for \mbox{three-} or four-body interactions. However, we are limited to simulating at most 11 atoms for the \mbox{three-} and four-body interactions with the $spp's'$ model; beyond that the number of states becomes too large (see Sec.~\ref{sec:counting} in the Appendix for details on how the states are counted). 

The limited number of atoms presents two challenges to studying the dynamics. First, we would like to be in the regime where the spatial boundaries or small number of atoms do not determine the evolution. This is a particular concern in the case of \mbox{three-} and four-body interactions, where a small system size significantly limits the number of possible triplets or quadruplets of atoms. For example, our results change meaningfully as we increase the number of included atoms from 10 to 11. In this sense, with only 11 atoms, the dynamics have not yet converged (see Appendix~\ref{sec:appendix-comparison}).

Second, it is not possible to probe the evolution at long enough times to draw conclusions about the thermodynamic fate of the system. The timescale on which we expect the finite size of the system to become relevant is given by the Heisenberg time $t_H\propto 1/\Delta E$. We estimate $\Delta E$ from the average energy level spacing. We find that, using the $spp's'$ model, $t_H$ is generally too short to numerically study thermalization. This problem is explored in detail by Panda \textit{et al.}~\cite{panda_can_2020}.

In order to achieve better convergence and to probe longer times, we have developed the $sps'$ model, in which we omit the $p'$ energy level. In removing an energy level, we sacrifice the ability to quantitatively model the populations of the more realistic energy levels. However, we retain the essential qualitative feature of the experimental system: field-tuned few-body dipole-dipole interactions alongside two-body hopping interactions. 

For the $sps'$ model, the scaling of the size of the Hamiltonian with the number of atoms is more favorable and we can include 12 to 14 atoms in our simulations (again, see Sec.~\ref{sec:counting} in the Appendix for details on how the states are counted). Since $t_H$ scales with the size of the system, we can reliably probe longer times and study thermalization with the $sps'$ model. We present more details on the Heisenberg times and convergence of the $spp's'$ and $sps'$ models in Appendix~\ref{sec:appendix-comparison}.

In Sec.~\ref{sec:sps-scar} we show that quantum many-body scar states arise in the $sps'$ model for the \mbox{three-} or four-body interactions, just as for the $spp's'$ model. In Sec.~\ref{sec:slow-therm-sps}, we explore the dynamics of the $sps'$ model at longer times. We find that, for a wide range of interaction energies and spatial disorders, thermalization is slow and the dynamics are nonergodic. 

\section{The $\boldsymbol{spp's'}$ Model\label{sec:spps}}

Rubidium Rydberg atoms in $np$ states with ${32<n<38}$ can resonantly exchange energy through a set of few-body dipole-dipole interactions~\cite{tretyakov_observation_2017,liu_time_2020}. An applied electric field tunes these interactions into resonance via the Stark effect. For example, Liu \textit{et al.}\ initially excited a sample of atoms in a magneto-optical trap to the $36p_{3/2,|m_j|=1/2}$ state, which we label as the $p$ state. The energetically nearby $36p_{3/2,|m_j|=3/2}$, or $p'$ state, was initially unpopulated. As shown in Fig.~\ref{fig:spps-model}, two $p$ atoms can resonantly exchange energy by the two-body field-tuned interaction
\begin{equation}
    p + p \leftrightarrow s + s',\label{eq:spps-two}
\end{equation}
where $s$ refers to the $37s$ state and $s'$ refers to the $36s$ state. 

At a slightly larger electric field, the two-body interaction is tuned out of resonance by the energy ${\delta=E_p - E_{p'}}$ as shown in Fig.~\ref{fig:spps-model}. A third $p$ atom can be recruited to account for this defect via the ${s + p \rightarrow p' + s}$ interaction so that one atom ends up in the $p'$ state. By further increasing the electric field, the energy defect can be set to ${\delta=2(E_p-E_{p'})}$ so that a fourth $p$ atom is needed to bring the interaction into resonance. These \mbox{three-} and four-body interactions are Borromean in nature and require all atoms to participate simultaneously~\cite{tretyakov_observation_2017}.

Along with the two-body field-tuned interaction given by Eq.~(\ref{eq:spps-two}), we can summarize the field-tuned interactions in the $spp's'$ model as
\begin{eqnarray}
    p + p + p &\leftrightarrow& s + s' + p'\label{eq:three-body-spps}\\
    p + p + p + p &\leftrightarrow& s + s' + p' + p'.
\end{eqnarray}
The hopping interactions are always present and are given by
\begin{eqnarray}
    p + s &\leftrightarrow& s + p\label{eq:spps-ar1}\\
    p + s' &\leftrightarrow& s' + p\\
    p' + s &\leftrightarrow& s + p'\\
    p' + s' &\leftrightarrow& s' + p'.\label{eq:spps-ar}
\end{eqnarray}


Since the field-tuned \mbox{two-}, \mbox{three-}, and four-body interactions are resonant at different electric fields, we simulate each case separately. Even though all three Hamiltonians include two-body hopping interactions, we shall refer to them as the two-body, three-body, or four-body Hamiltonian operators, in reference to the field-tuned interaction. For example, we can write the three-body Hamiltonian operator as
\begin{multline}
  \hat{H}_{3(spp's')}=\sum_{i\ne j\ne k}\left(\hat{\sigma}_{ps}^i \hat{\sigma}_{ps'}^j \hat{\sigma}_{pp'}^k +\mathrm{H.c.} \right)V^{ijk}_3 \\
    + \sum_{i\ne j} \Bigl( \vphantom{\beta^2\hat{\sigma}_{s'p'}^j} \mu^2\hat{\sigma}_{ps}^i\hat{\sigma}_{sp}^j  + \nu^2\hat{\sigma}_{ps'}^i\hat{\sigma}_{s'p}^j \\+  \alpha^2\hat{\sigma}_{p's}^i\hat{\sigma}_{sp'}^j +  \beta^2\hat{\sigma}_{p's'}^i\hat{\sigma}_{s'p'}^j \Bigr)\frac{1}{R_{ij}^3},\label{eq:ham3-spps}
\end{multline}
where $\hat{\sigma}^i_{ps}$ is an operator that takes the $i^{th}$ atom from the $p$ to the $s$ state, the sums are performed over distinct triplets or pairs of atoms, and $\mu$, $\nu$, $\alpha$, and $\beta$ are the dipole moments coupling $p\leftrightarrow s$, $p\leftrightarrow s'$, $p'\leftrightarrow s$, and $p' \leftrightarrow s'$, respectively. The first sum represents the field-tuned three-body interactions, while the second sum represents the two-body hopping interactions. We assume that we are at the resonant electric field for the three-body interaction and ignore off-resonant terms like the two-body hopping interaction $p+s\leftrightarrow s'+p$ or the field-tuned \mbox{two-} and four-body interactions.

The factor $V_3^{ijk}$ in Eq.~(\ref{eq:ham3-spps}) is calculated by summing over all possible paths from the initial $\ket{ppp}$ state to the final $\ket{ss'p'}$ state, having adiabatically eliminated the detuned intermediate state. Ignoring the angular dependence, there are two possible paths so that 
\begin{multline}
V^{ijk}_3 = \frac{1}{\delta}\Bigl(\bra{ss'p'}\hat{\rho}^{jk}\ket{sps'}\bra{sps'}\hat{\rho}^{ik}\ket{ppp}\\ +\bra{ss'p'}\hat{\rho}^{ik}\ket{ps's}\bra{ps's}\hat{\rho}^{jk}\ket{ppp}\Bigr),
\end{multline}
where $\ket{sps'}$, for example, represents the product state $\ket{s}_i\ket{p}_j\ket{s'}_k$, $\hat{\rho}^{ij}$ is the dipole-dipole operator coupling atoms $i$ and $j$, and $\delta$ is the detuning of the intermediate steps. This yields
\begin{equation}
V^{ijk}_3 = \frac{\mu\nu(\beta\nu+\alpha\mu)}{\delta R_{jk}^3 R_{ik}^3}\label{eq:spps-H3}.
\end{equation}
The \mbox{two-} and four-body Hamiltonians are similar and are shown in Appendix~\ref{sec:app-ham}.

The dependence on the distance between atoms in Eq.~(\ref{eq:spps-H3}) results from summing over all possible paths from the initial state to the final state. While the two-body field-tuned resonant and hopping matrix elements are proportional to $R^{-3}$, the field-tuned resonant matrix elements of the \mbox{three-} and four-body interactions effectively scale as $R^{-6}$ and $R^{-9}$, respectively.

In all of our simulations, we arrange the atoms in a one-dimensional array, with the spacing between atoms $d$ chosen to be experimentally realistic. We vary this spacing to adjust the typical strength of the interactions. Results are presented using atomic units for energy and, in order to compare the different few-body interactions, we use a natural time unit. The time unit is the reciprocal of the matrix element for $N$ nearest neighbors interacting via the $N$-body interaction that are separated by the nominal spacing for that simulation run. 

The values of the dipole moments are calculated numerically at the resonant electric field~\cite{zimmerman_stark_1979}. For example, we have
\begin{equation}
 \mu=\bra{37s}Ez\ket{36p_{3/2,|m_j|=1/2}},
\end{equation}
where $E$ is the applied electric field pointing in the $z$-direction, and $\ket{37s}$ and $\ket{36p_{3/2,|m_j|=1/2}}$ are the Stark states adiabatically connected to the zero-field $37s$ and $36p_{3/2,|m_j|=1/2}$ states. All of the dipole moments are about 700~$ea_0$. The value of $\delta$ is determined from the numerically calculated energies of the $36p_{3/2,|m_j|=1/2}$ and $36p_{3/2,|m_j|=3/2}$ Stark states. 

All simulations are run either on a local supercomputer with four Nvidia A100 graphical processing units or on similar NSF ACCESS resources~\cite{boerner_access_2023}. Linear algebra operations are performed with the MAGMA software package~\cite{tomov_dense_2010,tomov_dense_2010a, dongarra_accelerating_2014}.

\subsection{Time evolution in the $\boldsymbol{spp's'}$ model}

\begin{figure}
    \centering
    \includegraphics{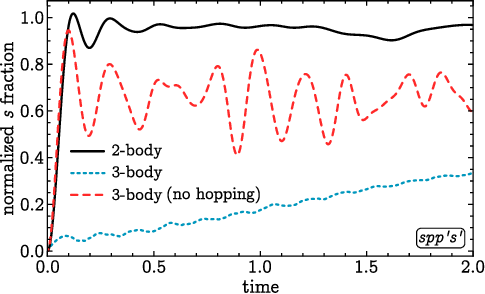}
    \caption{The fraction of atoms excited to the upper $s$ state normalized to the predicted saturation level for the $spp's'$ model as a function of time in natural time units. The two-body case is shown in solid black and is for 12 atoms in an array with a 50~$\mu$m spacing, using the Hamiltonian of Eq.~(\ref{eq:ham2-sps}). The three-body case is shown in dotted blue for 11 atoms in an array with a 9~$\mu$m spacing, using the Hamiltonian of Eq.~(\ref{eq:ham3-spps}). For these spacings, nearest neighbor pairs in the two-body case have a similar field-tuned matrix element as nearest neighbor triplets in the three-body case. The dashed red line shows the result when the hopping interaction is artificially disabled in the simulation for the three-body interaction.}
    \label{fig:spps-order}
\end{figure}

We start with an example that displays the typical time evolution of the few-body interactions. We compare the time evolution of the two-body interaction with $d=50$~$\mu$m to the three-body interaction with $d=9$~$\mu$m, using the $spp's'$ model. For these array spacings, the field-tuned two-body matrix elements are nearly the same as the field-tuned three-body matrix elements. The initial state is $\ket{\Psi(t=0)}={\ket{pp\cdots p}}$ and interactions are quenched into resonance at $t=0$. Figure~\ref{fig:spps-order} shows the normalized fraction of atoms that end up in the $s$ state as a function of time, with the data displayed as a fraction of the expected saturation level. The saturation level is calculated under the assumption that the ETH is obeyed and all many-body eigenstates are thermodynamically equivalent. The two-body $s$ saturation level is 0.326, the three-body is 0.241, and the four-body is 0.193. Even though the matrix elements are similar, the two-body $s$ fraction saturates quickly compared to the three-body $s$ fraction, which eventually approaches saturation after $t=10$.

Given that the matrix elements are similar, what causes the significantly slower evolution in the three-body case? An essential difference between the \mbox{two-} and three-body cases shown in Fig.~\ref{fig:spps-order} is the relationship between the always-resonant hopping and the field-tuned interactions. In the two-body case, the field-tuned and hopping matrix elements are approximately equal and both scale as $R^{-3}$. However, in the three-body case, the field-tuned matrix element scales as $R^{-6}$ and is thus significantly smaller than the hopping matrix element, which still scales as $R^{-3}$. 

We can numerically test the effect of the hopping interactions in the three-body case by artificially disabling them in our simulation. The resulting time evolution of the three-body dynamics, shown in Fig.~\ref{fig:spps-order}, is dramatically different from the unmodified three-body case. Since the stronger hopping interactions have been removed, the lower frequency Rabi oscillations of the three-body interaction are now visible. By comparing the first oscillation in the two-body case to the first oscillation in the three-body case without hopping, it is evident that the \mbox{two-} and three-body matrix elements have the same magnitude. The greater visibility of the three-body Rabi oscillations compared to the two-body case is due to the relatively smaller number of possible triplets in the 11-atom three-body case compared to the number of possible pairs in the 12-atom two-body case.

Furthermore, we see that the artificially modified three-body normalized $s$ fraction in Fig.~\ref{fig:spps-order} increases just as quickly as the two-body and reaches a higher level than the unmodified three-body. We can conclude that the always-resonant hopping interaction is an important factor in the relatively slow dynamics of the unmodified three-body case. One might not expect the presence of hopping interactions to slow the dynamics; however, localization has been found previously in systems with long-range hopping~\cite{celardo_shielding_2016,santos_cooperative_2016,deng_duality_2018,nosov_correlationinduced_2019}.

\subsection{The $\boldsymbol{spp's'}$ model with only two-body interactions}

To isolate the effect of the hopping interactions from the three-body nature of the field-tuned energy exchange, we can construct a model with only two-body interactions that retains the essential feature of relatively stronger hopping interactions. In the results of Fig.~\ref{fig:spps-order}, we disabled the hopping interactions to show that the \mbox{two-} and three-body dynamics become more similar. However, when we tune from the two-body to the three-body resonant field, the hopping matrix elements remain constant while the field-tuned matrix elements become smaller. Thus, in this new model we artificially adjust the strength of the field-tuned interactions relative to the hopping interactions. The two-body field-tuned matrix element is modified to be $\kappa \mu\nu/R^3$, where $0 < \kappa \le 1$ and $\mu$ and $\nu$ are the dipole moments coupling $p\leftrightarrow s$ and $p \leftrightarrow s'$, respectively. The hopping matrix elements are left unmodified. The parameter $\kappa$ allows us to weaken the two-body field-tuned matrix element relative to the hopping matrix elements, which is analogous to the three-body case.

\begin{figure}[!ht]
    \centering
    \includegraphics{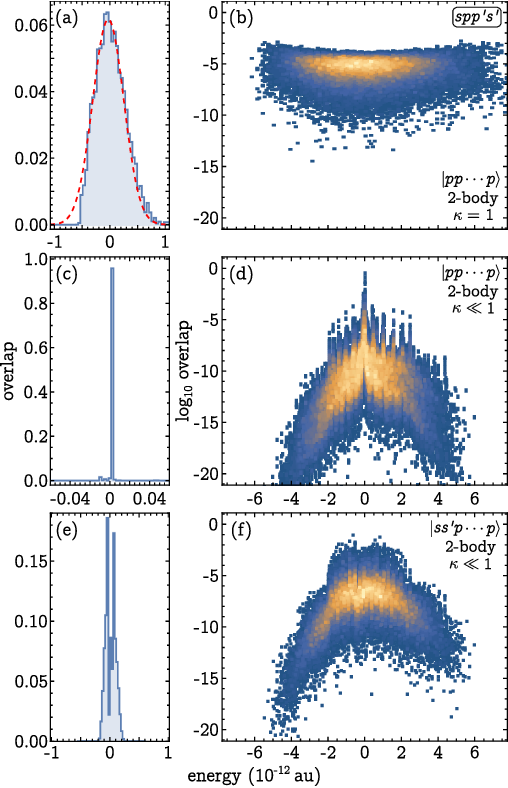}
    \caption{Overlap of the many-body energy eigenstates with the initial state for 12 atoms with a 50~$\mu$m spacing undergoing two-body interactions. These results use the two-body $spp's'$ Hamiltonian of Eq.~(\ref{eq:ham2-spps}), with the field-tuned matrix element modified to be $\kappa\mu\nu/R^3$. In (a)-(d), the initial state is $\ket{pp\cdots p}$. In (a), (c), and (e), the LDOS is shown with the overlap binned by energy. In (b), (d), and (f), the $\log_{10}$ of the overlap with the initial state as a function of energy is shown for each eigenstate as a scatter plot, with lighter colors indicating a higher density of points. In (a) and (b), $\kappa=1$ so that the field-tuned and hopping matrix elements are approximately equal. The LDOS in (a) is nearly Gaussian, as shown by the dashed red fit, and many eigenstates in (b) have similar overlap with the initial state. In (c) and (d), the field-tuned interaction matrix elements have been decreased by a factor of 60, a reasonable value for the three-body case. In (d), quantum many-body scar states emerge and have large overlap in narrow band near zero energy. In (e) and (f), the value of $\kappa$ is the same as in (c) and (d) but we start in the alternative initial state $\ket{ss'p\cdots p}$. In (f), no quantum many-body scar states are visible. }
    \label{fig:ldos}
\end{figure}

We can gain insight by examining the many-body eigenstates. The local density of states (LDOS) is the overlap of each eigenstate with the initial state, $\ket{\Psi(t=0)}=\ket{pp\cdot \cdot \cdot p}$, as a function of energy, given by
\begin{equation}
    \rho_0(E)=\sum_i \lvert \braket{\psi_i|\Psi(t=0)}\rvert^2\delta(E-E_i),
\end{equation}
where $E_i$ are the energy eigenvalues, $\ket{\psi_i}$ are the energy eigenstates, and the delta function selects the energy. For the unmodified case where  $\kappa=1$, Fig.~\ref{fig:ldos}(a) shows the LDOS, which is binned by energy. The LDOS fits well to a Gaussian which implies rapid thermalization~\cite{tavora_powerlaw_2017}.

We also graph $\log_{10}$ of the overlap of each eigenstate with the initial state as function of energy on a scatter plot as shown in Fig.~\ref{fig:ldos}(b). For the unmodified two-body case where $\kappa = 1$, no eigenstate or set of eigenstates dominates and all of the overlaps are smaller than 10$^{-2}$. The initial state is delocalized and ergodically spread across the energy eigenbasis.

Figures~\ref{fig:ldos}(c) and~\ref{fig:ldos}(d) show the results when $\kappa \ll 1$. Here, the field-tuned matrix elements are about 60 times smaller than the hopping matrix elements, which is an experimentally reasonable ratio for the three-body interaction. Even though the field-tuned interactions are so much weaker, they remain important for the dynamics. This is because $\ket{\Psi(t=0)}=\ket{pp\cdots p}$ so that only field-tuned interactions are initially possible. The hopping interactions of Eqs.~(\ref{eq:spps-ar1})-(\ref{eq:spps-ar}) cannot happen until a field-tuned interaction produces $s$ and $s'$ atoms. In this case, the LDOS is extremely narrow compared to the unmodified case in Fig.~\ref{fig:ldos}(a) (note the zoomed in horizontal scale in Fig.~\ref{fig:ldos}(c)). The initial state has significant overlap with only a few eigenstates, possibly indicating non-ergodic behavior~\cite{tavora_powerlaw_2017}.

In the scatter plot of Fig.~\ref{fig:ldos}(d), there are a few eigenstates near zero energy with large overlap with the initial state. There are also eigenstates, in bands of energies to the left and right of zero, that have atypically high overlap for that energy. These quantum many-body scar states have recently attracted significant interest~\cite{turner_weak_2018,turner_quantum_2018,moudgalya_entanglement_2018,choi_emergent_2019,lin_exact_2019,lin_quantum_2020,ho_periodic_2019,bluvstein_controlling_2021,serbyn_quantum_2021,langlett_rainbow_2022,schindler_exact_2022,iversen_tower_2023,evrard_quantum_2024,evrard_quantum_2024a} and have been proposed as the mechanism behind observed revivals in a Rydberg atom lattice~\cite{bernien_probing_2017a,turner_weak_2018}. The quantum many-body scar states fail to satisfy the ETH and form a subset of eigenstates that have strong overlap with the initial state or other product states. The subsequent time evolution can feature revivals as the system oscillates between the initial state and this small subset. 

The high overlap of the initial state with the scar states presumably contributes significantly to the slow thermalization visible for the three-body interaction in Fig.~\ref{fig:spps-order}. We can test this hypothesis by using a different initial state. Instead of the initial state $\ket{pp\cdots p}$ used in Fig.~\ref{fig:ldos}(a)-(d), consider the initial state $\ket{ss'p\cdots p}$, in which one nearest-neighbor pair of atoms starts in $ss'$. The LDOS shown in Fig.~\ref{fig:ldos}(e), while not Gaussian, is only slightly narrower than in the unmodified case of Fig.~\ref{fig:ldos}(a). The scatter plot of Fig.~\ref{fig:ldos}(f) does not show evidence of scar states and the dynamics are much faster in this case.

This establishes two conditions under which quantum many-body scars become relevant to the dynamics. First, the field-tuned interactions must be much weaker than the hopping interactions. Second, the stronger hopping interactions are not possible in the initial state. We note that these are similar conditions under which quantum many-body scars arose in the Rydberg experiment of Bernien \textit{et al.}~\cite{bernien_probing_2017a,turner_weak_2018}. In that experiment, the weaker interaction coupled the ground state to the Rydberg state while the strong interaction was a van der Waals coupling between nearest-neighbor Rydberg states. The initial state $\ket{\mathbb{Z}_2}$, where Rydberg and ground state atoms alternate along the chain, has high overlap with quantum many-body scar states. Furthermore, since there are no nearest-neighbor Rydberg states in the initial state $\ket{\mathbb{Z}_2}$, the stronger van der Waals interaction is not possible.



\subsection{Quantum many-body scar states in the full $\boldsymbol{spp's'}$ model}

\begin{figure}
    \centering
    \includegraphics{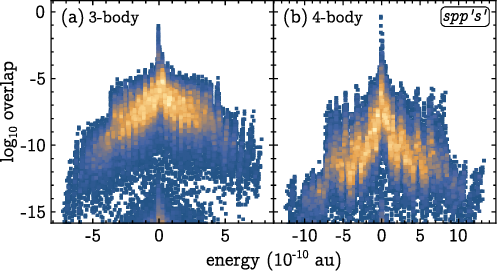}
    \caption{Scatter plots of $\log_{10}$ of the overlap of each of the eigenstates with the initial state for (a) \mbox{three-} and (b) four-body interactions in a linear array with 11 atoms using the $spp's'$ model. The initial state is $\ket{pp\cdots p}$. The spacings are 9~$\mu$m and 7~$\mu$m, respectively, for which the field-tuned matrix elements are similar in magnitude. In each case, a narrow band of quantum many-body scar states near zero energy with anomalously high overlaps is visible. In (a) we use the three-body $spp's'$ Hamiltonian of Eq.~(\ref{eq:ham3-spps}) and in (b) we use the four-body $spp's'$ Hamiltonian of Eq.~(\ref{eq:ham4-spps}).}
    \label{fig:3-4-spps-scatter}
\end{figure}

Leaving our model that only includes two-body interactions, we can visualize the eigenstates for the \mbox{three-} and four-body interactions in the same way. A scatter plot of the overlap as a function of energy for the three-body example of Fig.~\ref{fig:spps-order}, with 11 atoms and $d=9$~$\mu$m, reveals similar quantum many-body scar states, as shown in Fig.~\ref{fig:3-4-spps-scatter}(a). In Fig.~\ref{fig:3-4-spps-scatter}(b), we show the results for a linear array with 11 atoms and a 7~$\mu$m spacing undergoing four-body field-tuned interactions, for which the scar states are even more apparent. The field-tuned matrix elements are similar in magnitude for these two cases, but the hopping interactions are significantly stronger for the more closely spaced atoms in the four-body simulation. The emergence of quantum many-body scar states in the \mbox{three-} and four-body interactions is the primary driver of their slow dynamics.

\section{The $\boldsymbol{sps'}$ model}\label{sec:sps-scar}

\begin{figure}
    \centering
    \includegraphics{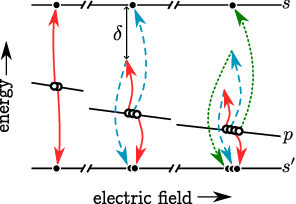}
    \caption{The \mbox{two-}, \mbox{three-}, and four-body interactions for the three-level $sps'$ model. The horizontal axis is the applied electric field while the vertical axis is energy. The pairs of solid red, dashed blue, and dotted green arrows represent individual detuned steps in the \mbox{three-} and four-body interactions. Open circles indicate initial states and closed circles indicate final states. While these circles are horizontally offset to allow space for the transition arrows, the individual resonant interactions occur at only one electric field. The detuning $\delta$ of the three-body interaction is indicated by the double-arrow at top-center. 
    These energy exchanges are described by Eqs.~(\ref{eq:two-body-sps})-(\ref{eq:ssar-sps}) and the Hamiltonians are given by Eq.~(\ref{eq:ham3-sps}), Eq.~(\ref{eq:ham2-sps}), and Eq.~(\ref{eq:ham4-sps}).}
    \label{fig:sps-model}
\end{figure}

When including the maximum number of atoms in the $spp's'$ model, the dynamics have not converged and the Heisenberg time is often too short to reliably simulate the dynamics at long times. To address these problems, we have developed a simplified $sps'$ model that involves only three energy levels: an initial state $p$, an upper final state $s$, and a lower final state $s'$. This model presents a similar set of field-tuned resonant \mbox{two-}, \mbox{three-}, and four-body interactions, as shown in Fig.~\ref{fig:sps-model}, given by
\begin{eqnarray}
    p + p &\leftrightarrow& s + s'\label{eq:two-body-sps}\\
    p + p + p &\leftrightarrow& s + s' + s'\label{eq:three-body-sps}\\
    p + p + p + p &\leftrightarrow& s + s' + s' + s'.\label{eq:four-body-sps}
\end{eqnarray}
The $sps'$ model retains the essential physical feature of the $spp's'$ model: three- and four-body field-tuned interactions alongside relatively stronger hopping interactions.
 
We have labeled the states of the $sps'$ model to parallel the $spp's'$ model for easy comparison. As explained below, the three- and four-body interactions in the $sps'$ model require a coupling between every pair of states. A transition between two $ns$ states is forbidden by angular momentum selection rules. However, dipole-dipole interactions have recently been observed been observed among Rydberg atoms excited to a state in the middle of the Stark manifold~\cite{opsahl_energy_2024}. These manifold states are broad superpositions of many zero-field angular momentum eigenstates even at relatively small electric fields. Thus, dipole transitions are allowed between nearly every pair of states in the manifold that satisfy the $m_j$ selection rules, even if forbidden at zero field~\cite{pelle_quasiforbidden_2016}. Three-body interactions that would be well-described by the $sps'$ model are likely to play a role in this system.



Regardless, our motivation for removing the $p'$ state is to create a simpler model that retains the essential dynamics of the $spp's'$ model. We have run the full set of simulations presented in Sec.~\ref{sec:sps-scar} and Sec.~\ref{sec:slow-therm-sps} both with and without the $s\leftrightarrow s'$ hopping interactions, and the results are essentially identical. Thus, there are three possible hopping interactions
\begin{eqnarray}
    p + s &\leftrightarrow& s + p\\
    p + s' &\leftrightarrow& s' + p.\\
    s + s' &\leftrightarrow& s' + s.\label{eq:ssar-sps}
\end{eqnarray}


Just as in the $spp's'$ model, the \mbox{three-} and four-body matrix elements for the $sps'$ model can be calculated perturbatively by considering a series of detuned two-body interactions. For example, in the three-body energy exchange of Eq.~(\ref{eq:three-body-sps}) shown in Fig.~\ref{fig:sps-model}, the initial two-body interaction $p+p\rightarrow s+s'$ is detuned by $E_p-E_{s'}$. The subsequent two-body interaction $p+s\rightarrow s+s'$ accounts for this detuning and brings the three-body interaction into resonance. Similarly, the four-body interaction is detuned by $2(E_p-E_{s'})$ and can be made resonant with two additional two-body hopping exchanges. The magnitude of $\delta$ in the $sps'$ model is larger than for the $spp's'$ model. However, we artificially set $\delta$ in the $sps'$ model to have the same value as for the $spp's'$ model so as to keep the matrix elements similar in magnitude.

The three-body Hamiltonian, for example, is
\begin{equation}
\begin{split}
    \hat{H}_{3(sps')}&=\sum_{i\ne j\ne k}\left(\hat{\sigma}_{ps}^i \hat{\sigma}_{ps'}^j \hat{\sigma}_{ps'}^k +\mathrm{H.c.} \right)V^{ijk}_3 \\
    +& \sum_{i\ne j} \left( \mu^2\hat{\sigma}_{ps}^i\hat{\sigma}_{sp}^j + \nu^2\hat{\sigma}_{ps'}^i\hat{\sigma}_{s'p}^j +  \xi^2\hat{\sigma}_{ss'}^i\hat{\sigma}_{s's}^j \right)\frac{1}{R_{ij}^3},
    \end{split}\label{eq:ham3-sps}
\end{equation}
where $\hat{\sigma}^i_{ps}$ is an operator that takes the $i^{th}$ atom from the $p$ to the $s$ state and the sums are performed over distinct triplets or pairs of atoms. We assume that we are at the resonant electric field. The second term represents the two-body hopping interactions while the first term is the field-tuned resonant three-body interaction with
\begin{equation}
V^{ijk}_3 = \frac{\xi \mu^2 \nu}{\delta R_{12}^3}\left(\frac{1}{R_{13}^3} + \frac{1}{R_{23}^3}\right)\label{eq:H3},  
\end{equation}
where $R_{ij}$ is the distance between atoms $i$ and $j$. The detuning from the two-body resonance is $\delta$ and $\mu$, $\nu$, and $\xi$ are the dipole moments coupling $p\rightarrow s$, $p\rightarrow s'$, and $s\rightarrow s'$, respectively. Again, the field-tuned resonant matrix elements of the \mbox{three-} and four-body interactions effectively scale as $R^{-6}$ and $R^{-9}$, respectively. The \mbox{two-} and four-body Hamiltonian operators are shown in Appendix~\ref{sec:app-ham}.

In Sec.~\ref{sec:spps}, we found that quantum many-body scar states arise in the $spp's'$ model when the matrix elements for the \mbox{three-} or four-body interactions were significantly smaller than the matrix elements for the hopping interactions. Since the $sps'$ model retains the essential physical characteristics of the $spp's'$ model, one expects that it will also present quantum many-body scar states. In fact, for a two-body interaction, the $sps'$ and $spp's'$ models are nearly identical. Thus, the results shown in Fig.~\ref{fig:ldos} also apply to the $sps'$ model. 

\begin{figure}
    \centering
    \includegraphics{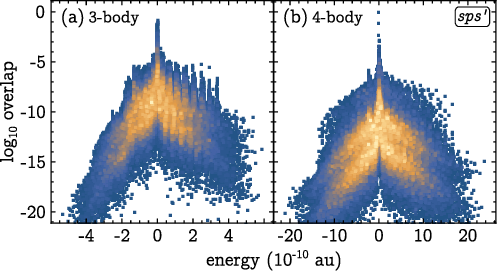}
    \caption{Scatter plots of $\log_{10}$ of the initial state overlap of the eigenstates for (a) three-body interactions with 13 atoms and (b) four-body interactions with 14 atoms in a linear array using the $sps'$ model. The initial state is $\ket{pp\cdots p}$. The spacings are 11~$\mu$m and 7~$\mu$m, respectively, for which the field-tuned matrix elements are similar in magnitude. In each case, a narrow band of quantum many-body scar states near zero energy with anomalously high overlaps is visible. In (a) we use the three-body $sps'$ Hamiltonian of Eq.~(\ref{eq:ham3-sps}) and in (b) we use the four-body $sps'$ Hamiltonian of Eq.~(\ref{eq:ham4-sps}).}
    \label{fig:3-4-ldos}
\end{figure}

We can also visualize the eigenstates for the \mbox{three-} and four-body interactions in the same way. A scatter plot of the $\log_{10}$ of the overlap of each eigenstate with the initial state is shown in Fig.~\ref{fig:3-4-ldos} for two cases with similar matrix elements in the $sps'$ model. Figure~\ref{fig:3-4-ldos}(a) shows the three-body interaction for a spacing of 11~$\mu$m and 13 atoms and Fig.~\ref{fig:3-4-ldos}(b) shows the four-body interaction for a spacing of 7~$\mu$m and 14 atoms. In both cases, a narrow band of scar states is visible near zero energy and the overall structure is quite similar to that of the $spp's'$ model in Fig.~\ref{fig:3-4-spps-scatter}.

\section{Slow thermalization in the $\boldsymbol{sps'}$ model}\label{sec:slow-therm-sps}

In order to numerically study thermalization or localization, it is necessary to simulate the system to sufficiently long times. The finite size of the system becomes relevant after the Heisenberg time, so that the dynamics are partly determined by the boundaries. Panda \textit{et al.}\ discuss the possibility that the time that it takes the system to thermalize might be longer than the Heisenberg time~\cite{panda_can_2020}. Since one does not know \textit{a priori} if this is the case, they argue that system sizes significantly larger than typically used are needed to simulate the dynamics at long times.

While we are limited to including 12 atoms for two-body interactions, 13 atoms for three-body interactions, and 14 atoms for four-body interactions, we find that $t_H$ is typically long enough to study thermalization in the $sps'$ model. When possible, we display the Heisenberg time on our plots. For more discussion on convergence of the models, see Appendix~\ref{sec:appendix-comparison}, and for details on how the states are counted, see Appendix~\ref{sec:counting}.

In order to study the thermalization of few-body dipole-dipole interactions among Rydberg atoms, we examine the following set of dynamical variables in our simulations. We calculate the fraction of atoms excited to the $s$ state as a function of time, which has the advantage of being straightforward to measure in experiment~\cite{gregoric_improving_2018, gregoric_perturbed_2020,liu_time_2020}. We also determine the fidelity as a function of time, the mean energy level spacing, and the entanglement entropy as a function of time. The latter three of these are used widely in numerical studies to distinguish between thermalizing and nonergodic systems. 


Disorder is added to our one-dimensional model by shifting the position of the $i^{th}$ atom by $w_i d$ along the array axis, where $w_i$ is a random number drawn from the uniform distribution $[-w,w]$. We simulate our system for five disorders ranging from $w=0.05$ to $w=0.45$ in increments of 0.10. For the three-body case, we examine seven linear array spacings ranging from $d=7$~$\mu$m to $d=13$~$\mu$m, and for the four-body case, five spacings from $d=4$~$\mu$m to $d=8$~$\mu$m. In both cases, the range of $d$ was chosen to demonstrate the entire range of behaviors while also being experimentally realizable. For example, at the Rydberg atom densities studied in~\cite{liu_time_2020}, typical values for the two-body matrix elements were a few hundreds of kHz while the \mbox{three-} and four-body matrix elements were a few kHz. In our simulations, from the largest to the smallest linear array spacings, the \mbox{three-} and four-body matrix elements range from a few kHz to a few hundred kHz. For each case we average over at least 100 samples. 

The archetypal model for studying thermalization is the disordered interacting spin chain, in which the spins are fixed on a one-dimensional lattice and the disorder is introduced via random on-site potentials. In such models, the disorder and the interaction strength can be adjusted independently. However, in our model, disorder and interaction strength are coupled, since perturbing the positions of the atoms necessarily changes the dipole-dipole matrix elements. 

In Sec.~\ref{sec:fidelity}, we examine the initial state fidelity and normalized $s$ fraction as a function of time. We analyze our results using a method developed by T\'{a}vora \textit{et al.}\ to predict thermalization based on the power-law decay of the fidelity~\cite{tavora_powerlaw_2017}. In Sec.~\ref{sec:mean-level-sps}, we estimate the mean energy level spacing, which has been used to study the transition between thermalization and localization~\cite{pal_manybody_2010,barlev_absence_2015,burin_manybody_2015}. In Sec.~\ref{sec:ee-sps}, we study the growth of the entanglement entropy. Finally, in Sec.~\ref{sec:summary}, we summarize our results.

\subsection{Fidelity and normalized $\boldsymbol{s}$ fraction\label{sec:fidelity}}

We first examine the long-time behavior of the fidelity. The survival probability or fidelity, $F(t)$, is defined as 
\begin{equation}
    F(t)=\lvert  \braket{\Psi(t=0)|\Psi(t)}\rvert^2,
\end{equation}
and it gives the probability of finding the system in its initial state $\ket{\Psi(t=0)}$ at some later time $t$. T\'{a}vora \textit{et al.}\ have developed a method for predicting the onset of thermalization in isolated many-body systems based on fits of the decay profile of $F(t)$ at long times~\cite{tavora_powerlaw_2017}. They apply their method to an interacting spin-$\frac{1}{2}$ model in a one-dimensional lattice. After a rapid exponential collapse, the fidelity is best described by a power-law decay $F(t)$~$\propto$~$t^{-\gamma}$ where the value of $\gamma$ is shown to predict the thermodynamic fate. T\'{a}vora \textit{et al.}\ find that $\gamma \geq 2$ indicates thermalization, $\gamma < 1$ indicates nonergodic evolution, and intermediate values require further analysis.


\begin{figure}
    \centering
    \includegraphics{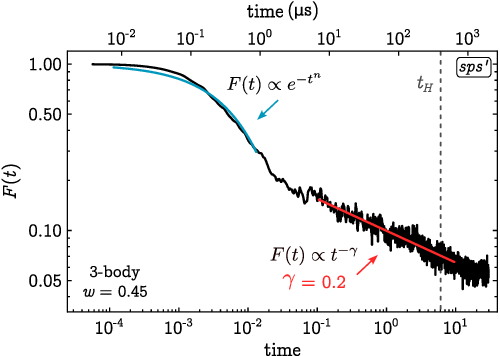}
    \caption{A log-log plot depicting a typical case of the fidelity $F(t)$ for a system undergoing three-body interactions for $d=9$~$\mu$m and ${w=0.45}$ using the $sps'$ model with the Hamiltonian of Eq.~(\ref{eq:ham3-sps}). The lower horizontal axis is time in natural time units and the upper is time in $\mu$s. The vertical dashed gray line represents the Heisenberg time. The blue region shows the initial rapid exponential collapse of $F(t)$. A power-law fit of $F(t)$ at long times is shown in red, with ${\gamma = 0.2}$ indicating slow thermalization.}
    \label{fig:gammafit}
\end{figure}

It is helpful to start by considering a typical example of the time evolution of the fidelity. Figure~\ref{fig:gammafit} shows three-body field-tuned and two-body hopping interactions with $w=$~0.45 and $d=$~9~$\mu$m. Initially, the fidelity collapses exponentially. This is followed by power-law decay at long times, in which a fit of $F(t)\propto t^{-\gamma}$ yields $\gamma=0.2$. 


The power-law decay in Fig.~\ref{fig:gammafit} begins to plateau around $F(t)=0.05$, near the end of the simulated time, after $t_H$. This plateau continues unchanged when we extend the simulated time. This is consistent with our estimate of the Heisenberg time, as one does not expect new dynamics after $t_H$. We can estimate the asymptotic value of $F(t)$ using the inverse participation ratio, or IPR, given by
\begin{equation}
    \mathrm{IPR}=\sum_i \lvert \braket{\psi_i|\Psi(t=0)}\rvert^4.\label{eq:IPR}
\end{equation}
The IPR essentially counts the number of eigenstates that significantly overlap the initial state, and $F(t)$ should therefore plateau around IPR$^{-1}$~\cite{tavora_inevitable_2016}. Indeed, for the simulation in Fig.~\ref{fig:gammafit}, we find IPR$^{-1}=0.046$, in agreement with the value read from the graph.

\begin{figure}
    \centering
    \includegraphics{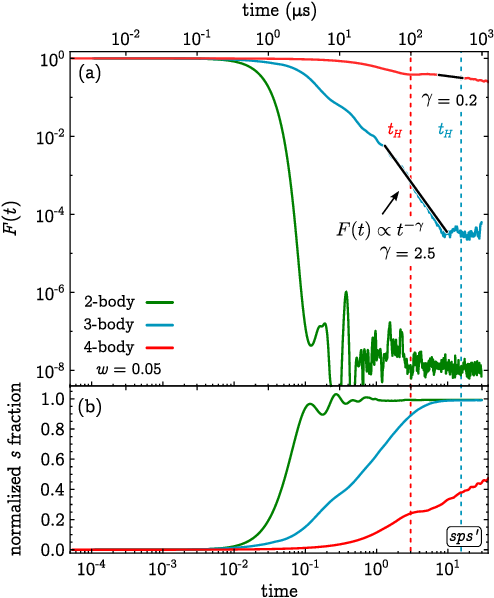}
    \caption{(a) Fidelity and (b) normalized $s$ fraction with low disorder ($w=0.05$) for \mbox{two-}, \mbox{three-}, and four-body field-tuned interactions in green, blue, and red, respectively. The results include hopping and are for the $sps'$ model using the Hamiltonians of Eqs.~(\ref{eq:ham2-sps}), (\ref{eq:ham3-sps}), and~(\ref{eq:ham4-sps}). The lower horizontal axis is in units of natural time. The respective array spacings of 32~$\mu$m, 8~$\mu$m and 6~$\mu$m mean that the field-tuned matrix elements are similar. The Heisenberg times are shown by the dashed lines in corresponding colors; the two-body $t_H$ is beyond the simulated time. In (a), the solid black lines represent fits to $F(t)\propto t^{-\gamma}$ at long times. $F(t)$ for the two-body case thermalizes too quickly to perform the fit. 
    } 
    \label{fig:2-3-4-fidplots-lowdis}
\end{figure}

\begin{figure}
    \centering
    \includegraphics{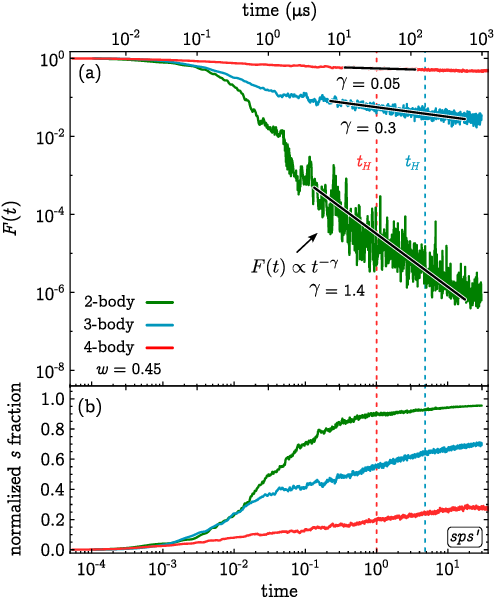}
    \caption{(a) Fidelity and (b) normalized $s$ fraction for the same systems as Fig.~\ref{fig:2-3-4-fidplots-lowdis}, but with high disorder ($w=0.45$). These results include hopping interactions and are for the $sps'$ model using the Hamiltonians of Eqs.~(\ref{eq:ham2-sps}), (\ref{eq:ham3-sps}), and~(\ref{eq:ham4-sps}) for the \mbox{two-}, \mbox{three-}, and four-body Hamiltonians, respectively. The Heisenberg times are shown by the dashed lines in corresponding colors; the two-body $t_H$ is beyond the simulated time. Fits of $F(t)\propto t^{-\gamma}$ at long times in (a) are represented by solid black lines. 
    } 
    \label{fig:2-3-4-fidplots-highdis}
\end{figure}

The simulated timescale in Fig.~\ref{fig:gammafit} extends beyond $t_H$, which is near 400~$\mu$s. This is beyond the typical lifetime of a Rydberg state; for example, the 36$p$ initial state studied in Liu \textit{et al.}~\cite{liu_time_2020} has a lifetime of about 40~$\mu$s~\cite{branden_radiative_2009,marcassa_measurement_2009}. Extending our simulated time in this way was necessary to obtain a good fit to the power-law decay region of $F(t)$. We note, however, that the onset of the power-law decay behavior is earlier than 10~$\mu$s and therefore accessible in experiment.

Figures~\ref{fig:2-3-4-fidplots-lowdis} and \ref{fig:2-3-4-fidplots-highdis} show plots of (a) the fidelity and (b) the normalized $s$ fraction as functions of time for \mbox{two-}, \mbox{three-} and four-body interactions with similar matrix elements. The plots in Fig.~\ref{fig:2-3-4-fidplots-lowdis} are all at low disorder, while the plots in Fig.~\ref{fig:2-3-4-fidplots-highdis} are at high disorder. The $s$ fraction is normalized relative to the expected saturation levels in the $sps'$ model, which are 0.326 for the two-body, 0.213 for the three-body, and 0.153 for the four-body. We also show the \mbox{three-} and four-body Heisenberg times; the two-body $t_H$ are beyond the simulated times.



For the two-body case, $F(t)$ in Fig.~\ref{fig:2-3-4-fidplots-lowdis}(a) rapidly collapses to nearly zero. This rapid thermalization is also evident in the two-body normalized $s$ fraction shown in Fig.~\ref{fig:2-3-4-fidplots-lowdis}(b), which saturates. A fit of the power-law decay parameter $\gamma$ for the high disorder two-body case in Fig.~\ref{fig:2-3-4-fidplots-highdis}(a) yields the intermediate value $\gamma=1.4$. Since $F(t)$ in this case decreases by six orders of magnitude while the normalized $s$ fraction in Fig.~\ref{fig:2-3-4-fidplots-highdis}(b) eventually saturates, our results suggest that this case thermalizes, albeit slightly more slowly. Indeed, we examined two-body interactions at all five disorders for $d=32$~$\mu$m, 39~$\mu$m, 47~$\mu$m, and 56~$\mu$m and found that all cases thermalize. 


The three-body cases shown in Fig.~\ref{fig:2-3-4-fidplots-lowdis} and Fig.~\ref{fig:2-3-4-fidplots-highdis} reveal slower dynamics. In Fig.~\ref{fig:2-3-4-fidplots-lowdis}(a), for low disorder, we find $\gamma=2.5$ and $F(t)$ approaching zero, predicting thermalization. The normalized $s$ fraction in Fig.~\ref{fig:2-3-4-fidplots-lowdis}(b) saturates, consistent with this conclusion. 
Increasing the disorder further slows the dynamics; the three-body case with $w=0.45$ in Fig.~\ref{fig:2-3-4-fidplots-highdis}(a) yields $\gamma=0.3$ and the normalized $s$ fraction fails to saturate before $t_H$.

We observe the slowest dynamics in the four-body cases of Fig.~\ref{fig:2-3-4-fidplots-lowdis} and Fig.~\ref{fig:2-3-4-fidplots-highdis}. The fidelity in both four-body cases fails to reach zero before the corresponding Heisenberg times. Fits to the long-time power-law decay yield $\gamma = 0.2$ and $\gamma = 0.05$ for $w=0.05$ and $w=0.45$ respectively, reflecting slow thermalization. For both four-body cases, the normalized $s$ fraction remains significantly lower than the \mbox{two-} and three-body cases.

While the field-tuned matrix elements for each of the cases shown in Figs.~\ref{fig:2-3-4-fidplots-lowdis} and \ref{fig:2-3-4-fidplots-highdis}  are of similar magnitude, the dynamics are considerably slower in the \mbox{three-} and four-body results. The results presented in Sec.~\ref{sec:spps} and Sec.~\ref{sec:sps-scar} show that quantum many-body scar states slow the evolution when there is zero disorder. We have confirmed that quantum many-body scar states persist for all disorders for the \mbox{three-} and four-body interactions. Increasing the disorder evidently further slows the dynamics.

Having examined some representative cases, we now fit the power-law decay for all simulated spacings and disorders. To obtain a robust fit, we would like to satisfy two conditions. First, there should be a sufficiently wide region of power-law decay before or around $t_H$, since this estimates when we expect the finite size of the system to play a role. Second, there should be enough power-law decay before $F(t)$ plateaus so that we can accurately determine the slope. We can usually satisfy both conditions, as in the case of Fig.~\ref{fig:gammafit} and most of the cases in Figs.~\ref{fig:2-3-4-fidplots-lowdis}(a) and~\ref{fig:2-3-4-fidplots-highdis}(a). In some cases, particularly for the four-body interactions, we are unable to satisfy the first condition but can still perform a fit satisfying the second condition. The four-body fit shown in Fig.~\ref{fig:2-3-4-fidplots-lowdis}(a) is an example where the estimated $t_H$ is earlier than the power-law decay. For more details, see Appendix~\ref{sec:appendix-comparison}.

\begin{figure}
    \centering
    \includegraphics{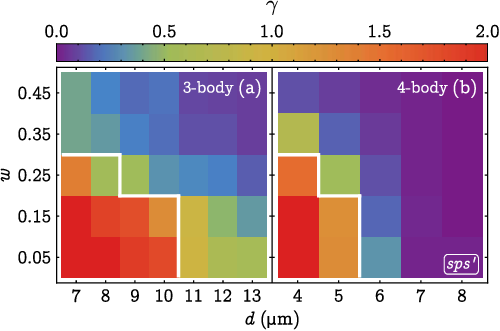}
    \caption{Intensity plots depicting the fits of the power-law decay parameter $\gamma$ as a function of array spacing $d$ and disorder $w$ in the $sps'$ model for (a) three-body interactions with the three-body Hamiltonian of Eq.~(\ref{eq:ham3-sps}), and (b) four-body interactions using the four-body Hamiltonian of Eq.~(\ref{eq:ham4-sps}). The parameter $\gamma$ is a result of the fit to the power-law decay of the initial state survival probability at long times. The color of the plot reflects the value of $\gamma$; red represents thermalization for $\gamma \geq 2$, whilst the cooler colors indicate nonergodic behavior for $0<\gamma<1$. For low enough disorder (small $w$) and high enough interaction energy (small $d$), $\gamma \geq 2$ and the system thermalizes. For larger disorders and smaller interaction energies, there are regions where $\gamma < 1$, indicating slow thermalization. The white lines differentiate the outcome of numeric fits of the entanglement entropy; the cases to the lower left of the white line demonstrate faster-than-logarithmic growth, whilst the remaining cases grow more slowly than logarithmically. }
    \label{fig:3-4-fidelitygrid}
\end{figure}

The summarized results for the fitted values of $\gamma$ are shown in Fig.~\ref{fig:3-4-fidelitygrid}. These intensity plots show the power-law decay parameter $\gamma$ as a function of $d$ and $w$ for the \mbox{three-} and four-body cases in Fig.~\ref{fig:3-4-fidelitygrid}(a) and Fig.~\ref{fig:3-4-fidelitygrid}(b) respectively. 

For the three-body case in Fig.~\ref{fig:3-4-fidelitygrid}(a), only the smallest values of $d$ with $w\leq 0.15$ are predicted to thermalize. Conversely, all spacings with $w\geq 0.35$ show slow thermalization. For larger array spacings where $d \geq 11$~$\mu$m, we observe slow thermalization regardless of disorder. This is expected for large enough $d$ as the hopping matrix elements are significantly greater than the field-tuned matrix elements in the three-body case, giving rise to quantum many-body scar states.

In the four-body case, the hopping matrix elements dominate even at small array spacings; we observe in Fig.~\ref{fig:3-4-fidelitygrid}(b) that thermalization is guaranteed only at $d=4$~$\mu$m with $w \leq 0.15$. Similar to the three-body case, we see that for $d\geq6$~$\mu$m, the system is nonergodic regardless of disorder. Since the four-body field-tuned matrix elements effectively scale as $R^{-9}$, they are significantly smaller than the hopping matrix elements at almost all spacings. The combined effect of scar states and disorder leads to $\gamma < 1$ for most cases and $\gamma < 0.1$ for many.

\subsection{Mean level spacing}\label{sec:mean-level-sps}

\begin{figure}
    \centering
    \includegraphics{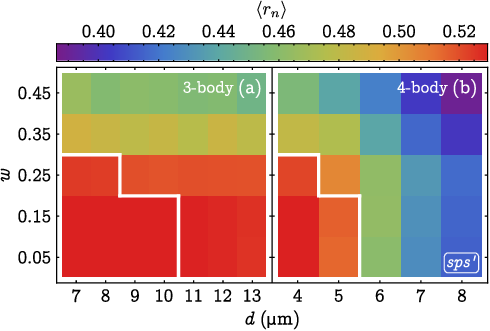}
    \caption{Computed mean level spacing $\langle r_n \rangle$ as a function of array spacing and disorder using the $sps'$ model for (a) three-body interactions and (b) four-body interactions using the three-body Hamiltonian of Eq.~(\ref{eq:ham3-sps}) and the four-body Hamiltonian of Eq.~(\ref{eq:ham4-sps}). Red reflects $\langle r_n \rangle$ consistent with a Wigner-Dyson distribution, whilst purple indicates a Poissonian distribution. In (a), we observe that $\langle r_n \rangle \approx 0.53$ for all values of $d$ when $w\leq 0.25$. This suggests a delocalized and thermalizing phase. In (b), this is true only for smaller values of $d$. Elsewhere, $\langle r_n \rangle < 0.53$, indicating possible slow thermalization and generally nonergodic behavior. In particular, in (b), we see that $\langle r_n \rangle$ approaches 0.386, and thus a Poissonian distribution, for larger $d$ and larger $w$. The regions to the lower left of the white lines are cases with faster-than-logarithmic growth of the EE.  }
    \label{fig:elss}
\end{figure}

We next consider the mean level spacing, a computational metric developed by Oganesyan and Huse in which the distribution of nearby energy levels predicts system behavior~\cite{oganesyan_localization_2007,pal_manybody_2010,burin_manybody_2015,barlev_absence_2015}. For adjacent many-body energy levels, the gap is $\delta_n = E_{n+1} - E_n \ge 0$ with eigenvalues $E_n$ listed in ascending order. The ratio of two consecutive gaps is defined as
\begin{equation}
    0 \le r_n = \frac{\mathrm{min}\{\delta_n,\delta_{n-1}\}}{\mathrm{max}\{\delta_n,\delta_{n-1}\}} \le 1.
\end{equation}
For nonergodic dynamics, Poissonian statistics best describe the energy spectrum with the average value $\braket{r_n}\approx 0.386.$ A Poissonian distribution for this uncorrelated phase is expected as nearby energy levels show no level repulsion and thus are nearly randomly distributed. Conversely, for the delocalized, thermalizing phase in which level repulsion is present, a Wigner-Dyson distribution in the Gaussian orthogonal ensemble best describes the energy spectrum, with  $\braket{r_n}\approx 0.5295$. We plot $\braket{r_n}$ as a function of array spacing and disorder for all \mbox{three-} and four-body cases in Fig.~\ref{fig:elss}(a) and Fig.~\ref{fig:elss}(b), respectively. 

For the three-body cases plotted in Fig.~\ref{fig:elss}(a), the values of $\langle r_n \rangle$ range from 0.531 to 0.450. Increasing disorder has the strongest effect on the level spacing statistics, as $\langle r_n \rangle$  begins to deviate from a Wigner-Dyson distribution for all $d$ with $w\geq0.35$. It is interesting to compare Fig.~\ref{fig:elss}(a) with Fig.~\ref{fig:3-4-fidelitygrid}(a). The mean level spacings are very close to Wigner-Dyson for all $w\le 0.25$. However, particularly for $w=0.25$, in many cases fits are consistent with $\gamma < 1$. This combination points to a delocalized phase that nonetheless thermalizes slowly. A similarly intermediate many-body critical phase was recently discovered and analyzed by Wang \textit{et al.}~\cite{wang_realization_2020,wang_manybody_2021}.

The four-body cases shown in Fig.~\ref{fig:elss}(b) are more consistent with the values of $\gamma$ shown in Fig.~\ref{fig:3-4-fidelitygrid}(b). When $d=7$~$\mu$m or 8~$\mu$m, we have both $\gamma < 0.1$ and $\braket{r_n}$ approaching Poissonian statistics.  Together, these indicate that an MBL phase is possible for the four-body interactions.

\subsection{Entanglement entropy}\label{sec:ee-sps}

Finally, we calculate the entanglement entropy (EE) as a function of time. We divide the linear array into two subsystems, the left and right halves, and calculate the bipartite Von Neumann EE, $S(t)$, which measures the degree of entanglement between the two subsystems. The EE is $S(t)=-\sum_i \lambda_i\ln\lambda_i$ where $\lambda_i$ are the eigenvalues of the reduced density matrix of one half of the system.

\begin{figure}
    \centering
    \includegraphics{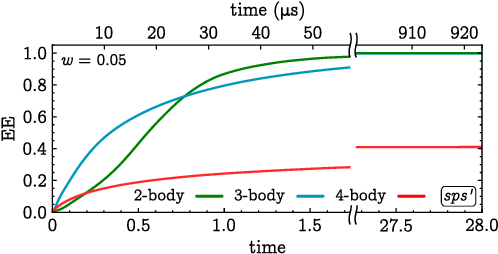}
    \caption{Entanglement entropy as a function of time for the same systems with low disorder $(w=0.05)$ shown in Fig.~\ref{fig:2-3-4-fidplots-lowdis}, in which the field-tuned matrix elements are similar in magnitude. The two-body case with an array spacing of 32~$\mu$m is shown in green, the three-body with 8~$\mu$m in blue, and the four-body with 6~$\mu$m in red. These results use the $sps'$ model with the \mbox{two-}, \mbox{three-}, and four-body Hamiltonians of Eq.~(\ref{eq:ham2-sps}), Eq.~(\ref{eq:ham3-sps}), and Eq.~(\ref{eq:ham4-sps}), respectively.    The time scale is broken in two with early times on the left, and times near the end of the simulation on the right. Following a short period of rapid initial growth, numeric fits determine that the \mbox{two-} and three-body cases thermalize as both grow faster than logarithmically and approach the maximum value. The four-body case is consistent with slower-than-logarithmic growth thermalizes slowly if at all. 
    }
    \label{fig:ee-lowdis}
\end{figure}

\begin{figure}
    \centering
    \includegraphics{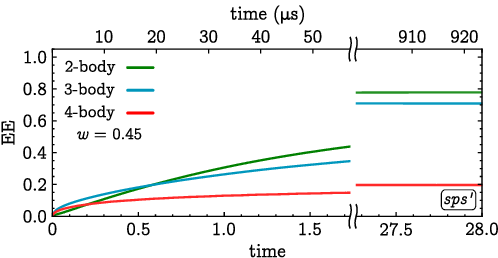}
    \caption{Entanglement entropy as a function of time for the same cases shown in Fig.~\ref{fig:2-3-4-fidplots-highdis}, or the same systems undergoing \mbox{two-}, \mbox{three-}, or four-body interactions for the $sps'$ model shown in Fig.~\ref{fig:ee-lowdis} but with high disorder $(w=0.45)$. Numeric fits determine that only the two-body case grows faster than logarithmically, whereas both the \mbox{three-} and four-body cases and display slower-than-logarithmic growth.}
    \label{fig:ee-highdis}
\end{figure}

The EE has been found to grow only logarithmically in the MBL phase for short-range systems~\cite{znidaric_manybody_2008,bardarson_unbounded_2012,kunimi_nonergodic_2021} and algebraically for the MBL phase in long-range systems~\cite{safavi-naini_quantum_2019,deng_universal_2020,lukin_probing_2019}, while it grows faster for thermalizing systems~\cite{kim_ballistic_2013}. The interactions $1/R^\beta$ are short-range if $\beta$ is greater than the dimension of the system, as in our case.  Figures~\ref{fig:ee-lowdis} and \ref{fig:ee-highdis} depict plots of the EE for the same cases shown in Figs.~\ref{fig:2-3-4-fidplots-lowdis} and \ref{fig:2-3-4-fidplots-highdis}, respectively. 

On the left portion of Figs.~\ref{fig:ee-lowdis} and \ref{fig:ee-highdis}, the entanglement entropies of the \mbox{three-} and four-body cases initially outpace that of the two-body. This is because the spacing in the \mbox{three-} and four-body cases is much smaller and therefore the hopping matrix elements are significantly larger. Once a single field-tuned interaction has entangled a small cluster of atoms, the resulting $s$ and $s'$ atoms are now able to hop and entangle more distant parts of the system. 

For all three cases, the EE first grows rapidly before settling into slower growth near $t=1$ in Figs.~\ref{fig:ee-lowdis} and \ref{fig:ee-highdis}. We fit the EE curves to determine their growth rate before they plateau at long times. Both the \mbox{two-} and three-body cases with $w=0.05$ in Fig.~\ref{fig:ee-lowdis} grow faster than logarithmically, indicating thermalization. This is consistent with the results of Fig.~\ref{fig:2-3-4-fidplots-lowdis}(a), as both have $\gamma > 2$. Only the four-body case with low disorder in Fig.~\ref{fig:ee-lowdis} grows slower than logarithmically, consistent with the corresponding fit of $\gamma < 1$ in Fig.~\ref{fig:2-3-4-fidplots-lowdis}(a).

When the disorder is increased to $w=0.45$, we find that the speed of the dynamics is further inhibited. The EE of both the \mbox{three-} and four-body cases in Fig.~\ref{fig:ee-highdis} slowly thermalize as both have slower-than-logarithmic growth with $\gamma < 1$ in Fig.~\ref{fig:2-3-4-fidplots-highdis}(a). Only the two-body case in Fig.~\ref{fig:ee-highdis} grows faster than logarithmically.

One might expect the long-time EE to saturate when the growth profile is consistent with thermalization. Both the \mbox{two-} and three-body cases with $w=0.05$ in Fig.~\ref{fig:ee-lowdis} indeed approach the maximum value, while the remaining cases, including that of the thermalizing two-body case with $w=0.45$, fail to do so. When disorder is added by varying the positions of the atoms within the lattice, we create close clusters of atoms. The atoms in these clusters interact strongly and partially decouple from the rest of the system, inhibiting the spread of entanglement~\cite{bardarson_unbounded_2012,lukin_probing_2019,abumwis_extended_2020}. 

In the two-body example of Fig.~\ref{fig:ee-highdis}, the growth of the EE can provide insight regarding the thermodynamic fate of a system with an intermediate value of $\gamma$.  This case, as shown in Fig.~\ref{fig:2-3-4-fidplots-highdis}(a), has $\gamma=1.4$. However, $F(t)$ collapses to nearly zero in Fig.~\ref{fig:2-3-4-fidplots-highdis}(a) and the normalized $s$ fraction approaches near saturation in Fig.~\ref{fig:2-3-4-fidplots-highdis}(b), both of which suggest thermalization. Furthermore, the EE in for this case grows faster than logarithmically. We conclude, therefore, that this case thermalizes.

We fit the growth of the EE for all of our simulations and summarize the results in Figs.~\ref{fig:3-4-fidelitygrid} and~\ref{fig:elss}. The cases to the lower left of the white boundary line grow faster than logarithmically, while the remaining cases above and to the right of the white line grow slower than logarithmically. These results are consistent with the results of $\gamma$ on  Fig.~\ref{fig:3-4-fidelitygrid}; when $\gamma \ge 2$, the EE grows faster than logarithmically, and when $\gamma < 1$, the growth is logarithmic or slower.

\subsection{Summary}\label{sec:summary}

Our results from the fidelity fits, the mean level spacing, and the EE broadly agree and predict slow thermalization for the \mbox{three-} and four-body interactions for cases with high disorder and larger array spacings. As $d$ increases, the ratio of field-tuned to hopping matrix elements decreases rapidly. Like that of the zero spatial disorder cases studied in Sec.~\ref{sec:spps} and Sec.~\ref{sec:sps-scar}, this results in quantum many-body scar states that slow the dynamics. The addition of disorder further slows the dynamics and, for high disorder, we observe very slow thermalization or even failure to thermalize. 

We observe that for all simulations in which $\gamma \geq 2$, the EE and $\langle r_n \rangle$ are consistent with thermalization. For the intermediate regime, with $1< \gamma < 2$, we find that most cases have faster-than-logarithmic EE growth and $\langle r_n \rangle$ is more consistent with Wigner-Dyson statistics. Thus, we conclude that these cases also thermalize, though more slowly than the $\gamma \geq 2$ regime. We find that for the majority of simulations that fit $\gamma < 1$, the EE and the mean level spacing statistics point toward slow thermalization or failure to thermalize entirely. These slowly thermalizing systems are excellent candidates for future studies of MBL and nonergodic behavior. For example, in the four-body case of Fig.~\ref{fig:ee-highdis} with $w=0.45$ and $d=6$~$\mu$m, we find $\gamma = 0.05$, $\langle r_n \rangle = 0.423$, and slower-than-logarithmic growth of the EE. 

\section{Conclusion}

We have presented simulated results for one-dimensional chains of Rydberg atoms exchanging energy via few-body dipole-dipole interactions. We analyzed the dynamics by studying the properties of the energy eigenstates, the decay of the initial state fidelity, the energy level spacing statistics, and the growth of entanglement entropy. While our model was motivated by ultracold Rydberg experiments, our results give broader insight into the rich dynamics of few-body interacting systems.

By studying a linear array of atoms with no spatial disorder, we found numerical evidence for quantum many-body scar states when the hopping matrix elements were significantly larger than the field-tuned matrix elements. This condition is met for a wide range of experimentally typical interatomic spacings for the \mbox{three-} and four-body interactions; thus, we expect that scar states play an important role in slowing their dynamics.

When spatial disorder is added to the linear array of atoms by randomly perturbing their positions, we find that thermalization is further inhibited. Power-law fits to the decay of the initial state fidelity, the mean energy level spacing ratio, and the slow growth of entanglement entropy all point to the possibility of many-body localization in our model. While we would need to include more atoms in simulation to draw a definitive conclusion about MBL, the three-body interactions in the $spp's'$ system are readily accessible in experiment.

Density-matrix renormalization group~\cite{schollwock_densitymatrix_2005} and matrix product state simulation methods~\cite{schollwock_densitymatrix_2011} can be effective for one-dimensional lattice systems if the EE spreads slowly. This condition is true for much of the parameter space we explored for \mbox{three-} and four-body interactions. If the positions of the atoms are fixed in our model, such methods may allow for larger system sizes to be simulated in the future. 

We plan to further explore the dynamics of these few-body interactions in experiment. For lower values of the principal quantum number, the three-body resonance is well-resolved from the two-body resonance so that they can be compared in experiment. Using an initial state that does not map so closely to a quantum many-body scar state may result in significantly altered dynamics for the three-body case. Additionally, varying the dimensionality of the excited sample of Rydberg atoms should produce different dynamics for the \mbox{two-} and three-body interactions.

\begin{acknowledgments}
We thank Ceren B. Dag for helpful discussion. 

This work was supported by the National Science Foundation under Grants No. 2011583 and No. 2011610, and S.E.S. is supported by the National Science Foundation Graduate Research Fellowship under Grant No. 2334429.

This work used the Delta system at the National Center for Supercomputing Applications through allocation PHY230142 from the Advanced Cyberinfrastructure Coordination Ecosystem: Services \& Support (ACCESS) program, which is supported by National Science Foundation grants \#2138259, \#2138286, \#2138307, \#2137603, and \#2138296.
\end{acknowledgments}

\section{Appendix}\label{sec:appendix}

In this appendix, we provide additional details of the $spp's'$ and $sps'$ models and compare them side-by-side. While the two models are not in exact numerical agreement, they share the same fundamental dynamics. In particular, both models have few-body dipole-dipole field-tuned resonant interactions that exist side-by-side with relatively stronger two-body hopping interactions. 

In Appendix~\ref{sec:app-ham}, we show the Hamiltonian operators for the \mbox{two-} and four-body interactions for both the $spp's'$ and $sps'$ models. The Hamiltonian operators for the three-body interactions were previously shown in Eq.~(\ref{eq:ham3-spps}) for the $spp's'$ model and Eq.~(\ref{eq:ham3-sps}) for the $sps'$ model. In Appendix~\ref{sec:counting}, we provide details on counting the number of states for each model, including the analytical formulas for the three-body cases. Finally, in Appendix~\ref{sec:appendix-comparison}, we compare the $spp's'$ and $sps'$ models for some generic cases.

\subsection{Hamiltonians for \mbox{two-} and four-body cases }\label{sec:app-ham}

The three-body dipole-dipole Hamiltonian operators for the $spp's'$ and $sps'$ models are given in the main text by Eq.~(\ref{eq:ham3-spps}) and Eq.~(\ref{eq:ham3-sps}), respectively. We write the two-body dipole-dipole Hamiltonian operator for the $sps'$ model as
\begin{equation}
\begin{split}
    \hat{H}_{2(sps')} &= \sum_{i \ne j} \left(\hat{\sigma}_{ps}^i \hat{\sigma}_{ps'}^j +\mathrm{H.c.} \right)\frac{\mu\nu}{R_{ij}^3}\\
    +& \sum_{i\ne j} \left( \mu^2\hat{\sigma}_{ps}^i\hat{\sigma}_{sp}^j + \nu^2\hat{\sigma}_{ps'}^i\hat{\sigma}_{s'p}^j \right)\frac{1}{R_{ij}^3},
    \end{split}\label{eq:ham2-sps}
\end{equation}
using the same notation as in Eq.~(\ref{eq:ham3-sps}), where the first term is the field-tuned interaction and the second term represents the hopping interactions. 
The four-body dipole-dipole Hamiltonian operator for the $sps'$ model, at the resonant electric field, is
\begin{equation}
\begin{split}
    \hat{H}_{4(sps')} &=\sum_{i\ne j\ne k \ne \ell}\left(\hat{\sigma}_{ps}^i \hat{\sigma}_{ps'}^j \hat{\sigma}_{ps'}^k \hat{\sigma}_{ps'}^\ell +\mathrm{H.c.} \right)V^{ijk\ell}_4 \\
    +& \sum_{i\ne j} \left( \mu^2\hat{\sigma}_{ps}^i\hat{\sigma}_{sp}^j + \nu^2\hat{\sigma}_{ps'}^i\hat{\sigma}_{s'p}^j +  \xi^2\hat{\sigma}_{ss'}^i\hat{\sigma}_{s's}^j \right)\frac{1}{R_{ij}^3}
    \end{split}\label{eq:ham4-sps}
\end{equation}
where the second term again represents the hopping interactions and the first term is the field-tuned interaction with
\begin{equation}
\begin{split}
    V^{ijk\ell}_4 =&\frac{\xi^2 \mu^3 \nu}{\delta^2}\left( \frac{1}{R_{i\ell}^3}\left(\frac{1}{R_{ij}^3R_{jk}^3} + \frac{1}{R_{ik}^3R_{jk}^3}\right)\right.\\ &+\frac{1}{R_{j\ell}^3}\left(\frac{1}{R_{ij}^3R_{ik}^3} + \frac{1}{R_{ik}^3R_{jk}^3}\right)\\ &+\left.\frac{1}{R_{j\ell}^3}\left(\frac{1}{R_{ij}^3R_{ik}^3} + \frac{1}{R_{ij}^3R_{jk}^3}\right) \right).
\end{split}
\end{equation}
Just as for the three-body interaction in Eq.~(\ref{eq:H3}), $V^{ijk\ell}_4$ results from summing over all possible paths from the initial state $\ket{pppp}$ to the final state $\ket{ss's's'}$. 

The $\xi^2$ terms in Eq.~(\ref{eq:ham2-sps}) and Eq.~(\ref{eq:ham4-sps}) represent the hopping interaction $s\leftrightarrow s'$. As explained in the main text, this transition is nominally forbidden by dipole selection rules. However, dipole-dipole energy exchange has recently been observed among atoms excited to Stark manifold states~\cite{opsahl_energy_2024}, and those interactions can involve dipole transitions among all energy levels. In particular, three-body interactions like those in the $sps'$ model likely play a role. In our simulations, we find that the $\xi^2$ term does not significantly affect the results.

The two-body dipole-dipole Hamiltonian operator for the $spp's'$ model is
\begin{equation}
\begin{split}
    \hat{H}_{2(spp's')} &= \sum_{i \ne j} \left(\hat{\sigma}_{ps}^i \hat{\sigma}_{ps'}^j +\mathrm{H.c.} \right)\frac{\mu\nu}{R_{ij}^3}\\
    +& \sum_{i\ne j} \left( \mu^2\hat{\sigma}_{ps}^i\hat{\sigma}_{sp}^j + \nu^2\hat{\sigma}_{ps'}^i\hat{\sigma}_{s'p}^j \right)\frac{1}{R_{ij}^3},
    \end{split}\label{eq:ham2-spps}
\end{equation}
using the same notation as in Eq.~(\ref{eq:spps-H3}), where the first term is the field-tuned interaction and the second term represents the hopping interactions. 

Finally, the four-body dipole-dipole Hamiltonian operator for the $spp's'$ model, when at the resonant electric field, is
  \begin{multline}
    \hat{H}_{4(spp's')}=\sum_{i\ne j\ne k \ne \ell}\left(\hat{\sigma}_{ps}^i \hat{\sigma}_{ps'}^j \hat{\sigma}_{pp'}^k \hat{\sigma}_{pp'}^\ell +\mathrm{H.c.} \right)V^{ijk\ell}_4 \\
    +\sum_{i\ne j} \Bigl( \vphantom{\beta^2\hat{\sigma}_{s'p'}^j} \mu^2\hat{\sigma}_{ps}^i\hat{\sigma}_{sp}^j  + \nu^2\hat{\sigma}_{ps'}^i\hat{\sigma}_{s'p}^j \\+  \alpha^2\hat{\sigma}_{p's}^i\hat{\sigma}_{sp'}^j +  \beta^2\hat{\sigma}_{p's'}^i\hat{\sigma}_{s'p'}^j \Bigr)\frac{1}{R_{ij}^3},\label{eq:ham4-spps}
    \end{multline}
where the second term is the hopping interactions and the first term is the field-tuned interaction where $V^{ijk\ell}_4$ is determined by a sum over all paths from the initial state $\ket{pppp}$ to the final state $\ket{ss'p'p'}$ and has the value
\begin{multline}
    \frac{\left(\alpha ^2 \mu ^3 \nu +\beta ^2 \mu  \nu^3 +2 \alpha  \beta  \mu ^2 \nu ^2\right)}{\delta^2  R_{k\ell}^3}\left(\frac{1}{R_{ik}^3 R_{j\ell}^3}+\frac{1}{R_{i\ell}^3 R_{jk}^3}\right). 
\end{multline}

\subsection{Counting states}\label{sec:counting}

We can count the states for the $sps'$ model in the following way, using the three-body interaction as an example. Consider $n$ total atoms and choose $m$ of them to interact, where $m$ is a multiple of three. Of the $m$ interacting atoms, choose one third of them to be in the $s$ state since the interaction takes $p+p+p\rightarrow s+s'+s'$. Multiplying these two combinations, $\binom{m}{n}\times\binom{m}{m/3}$, yields the total number of states when $m$ atoms interact. All that remains is to sum over all possible values of $m$ so that
\begin{equation}
N_{(sps')}=\sum_{m=0,3,...}^n\binom{n}{m}\binom{m}{m/3},  \label{eq:count-sps}  
\end{equation}
where $N_{(sps')}$ is the total number of states. The calculations for \mbox{two-} and four-body interactions proceed similarly.

The state counting for the $spp's'$ model is not too different. Again using the three-body interaction as an example, we choose $m$ interacting atoms and $m/3$ of them to be in the $s$ state. However, now we must choose $m/3$ of the remaining $2m/3$ atoms to be in the $p'$ state since $p+p+p\rightarrow s+p'+s'$. This yields
\begin{equation}
N_{(spp's')}=\sum_{m=0,3,...}^n\binom{n}{m}\binom{m}{m/3}\binom{2m/3}{m/3}   \label{eq:count-spps} 
\end{equation}
for the number of states.

In practice, it is easiest to generate the list of states numerically. This is accomplished by computationally listing all possible combinations of the energy levels for a given number of atoms. States that do not conserve energy are removed from the list. The number of states generated in this way agrees with the calculations in Eq.~(\ref{eq:count-sps}) and Eq.~(\ref{eq:count-spps}).

\begin{figure}
    \centering
    \includegraphics{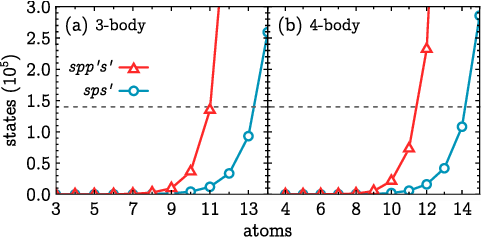}
    \caption{Number of states as a function of the number of included atoms for the $spp's'$ model, shown with red triangles, and the $sps'$ model, shown with blue circles. The three-body interaction is shown in (a) and the four-body interaction in (b). The dashed gray horizontal line is the maximum number of states that we can simulate on our hardware. }
    \label{fig:states}
\end{figure}

Figure~\ref{fig:states}(a) compares the number of states $N$ for the two models as a function of the number of atoms included in the simulation for the \mbox{three-} and four-body interactions. On our hardware, using exact diagonalization, we can simulate up to about $140\,000$ states. The $spp's'$ model only allows us to include 11 atoms for the \mbox{three-} and four-body interactions. However, we can include more atoms using the $sps'$ model: 13 atoms for the three-body interaction and 14 atoms for the four-body interaction.

\subsection{Comparison of the $\boldsymbol{sps'}$ and $\boldsymbol{spp's'}$ models}\label{sec:appendix-comparison}

The $spp's'$ model is defined by the energy level diagram of Fig.~\ref{fig:spps-model}, the energy exchanges of Eqs.~(\ref{eq:spps-two})-(\ref{eq:spps-ar}), and the Hamiltonians of Eqs.~(\ref{eq:ham2-spps}), (\ref{eq:ham3-spps}), and (\ref{eq:ham4-spps}). It includes all four relevant energy levels for the experiments described in Liu \textit{et al.}, where it was shown to provide accurate numerical predictions for the $s$ fraction at short times~\cite{liu_time_2020}. Current hardware limits us to including at most 12 atoms for the two-body interaction or 11 atoms for the \mbox{three-} and four-body interactions due to the size of the Hilbert space.

The $sps'$ model is defined by the energy level diagram of Fig.~\ref{fig:sps-model}, the energy exchanges of Eqs.~(\ref{eq:two-body-sps})-(\ref{eq:ssar-sps}), and the Hamiltonians of Eqs.~(\ref{eq:ham2-sps}), (\ref{eq:ham3-sps}), and (\ref{eq:ham4-sps}). In the $sps'$ model, we have dropped the $p'$ energy level, which reduces the size of the Hilbert space and therefore allows us to include more atoms: 13 atoms for the three-body case and 14 atoms for the four-body case. 

\begin{figure}
    \centering
    \includegraphics{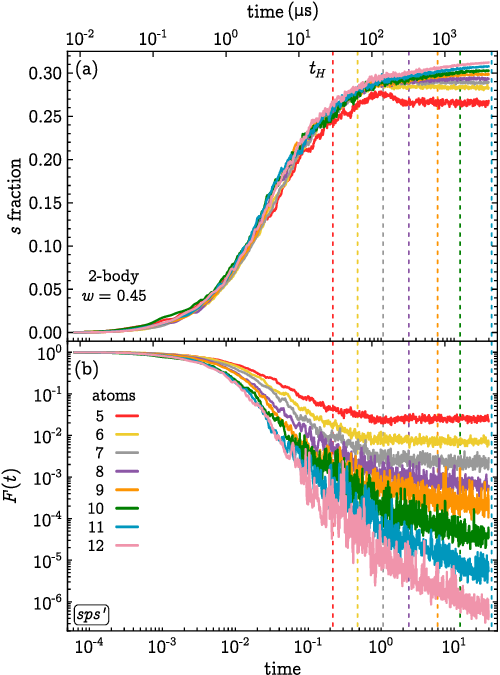}
    \caption{(a) $s$ fraction and (b) fidelity as a function of natural time (lower axis) and real time in $\mu$s (upper axis) for two-body field-tuned interactions in the $sps'$ model, using the Hamiltonian of Eq.~(\ref{eq:ham2-sps}). Each curve is colored for the number of atoms included in the simulation, as indicated by the legend. The vertical dashed lines show the corresponding Heisenberg time for each case. The $s$ fraction in (a) suggests that the results have converged, as all curves approach the expected saturation value of 0.326. The $F(t)$ curves in (b) do not converge. Rather, since the system thermalizes, each curve should approach the asymptotic value given by the reciprocal of the Hilbert space size, as discussed in the main text. More importantly, the slopes of the long-time power-law decay \textit{do} converge. In both the 11 and 12 atom cases, a fit to the slope of the long-time power-law decay $F(t)\propto t^{-\gamma}$ yields $\gamma=1.4$. The results for the two-body $spp's'$ model are essentially the same.}
    \label{fig:sps-2-body-conv}
\end{figure}

Since we have eliminated a relevant energy level, we no longer expect to accurately predict the state fractions as a function of time. However, we do retain the essential physical features of the more realistic $spp's'$ model: the few-body nature of the interactions and the differential scaling with interatomic distance of the hopping versus the field-tuned interactions. To the extent that these features determine the dynamics, and particularly the long-time thermodynamic evolution, we expect that results from the $sps'$ model should be generally applicable to the experimental $spp's'$ system and other similar systems. The comparisons below support this assertion. For each of these comparisons, we focus on the maximum disorder of $w=0.45$ as that represents the worst-case scenario in terms of the shortness of the Heisenberg times.

We start by examining the results for the field-tuned two-body interactions. In this case, the $spp's'$ and $sps'$ models are nearly identical. The only difference is the presence of the $\xi^2$ term in the Hamiltonians of Eq.~(\ref{eq:ham2-sps}) as compared to Eq.~(\ref{eq:ham2-spps}). This term represents the $s+s' \leftrightarrow s'+s$ hopping interaction, which does not exist in the $spp's'$ model. In our tests, this term leads to only small differences in the results. 

Figure~\ref{fig:sps-2-body-conv} shows the fidelity and the $s$ state fraction as a function of time for two-body interactions for the $sps'$ model for $w=0.45$ and $d=40$~$\mu$m. Examining the $s$ fraction in Fig.~\ref{fig:sps-2-body-conv}(a), it is plausible that the results have nearly converged, as most cases approach the expected saturation level of 0.326 with only slight differences at long times. Note that Fig.~\ref{fig:sps-2-body-conv}(a) shows the raw $s$ fraction, as opposed to the normalized $s$ fraction shown in earlier figures when comparing \mbox{two-}, \mbox{three-}, and four-body interactions.

The fidelity shown in Fig.~\ref{fig:sps-2-body-conv}(b) reveals more interesting trends as the number of included atoms increases. First, note that beyond $t_H$ the curves tend to flatten. This is because the $t_H$ gives the time scale on which we expect the finite size of the system to lead to periodic behavior; beyond that time no dynamical evolution is possible and we expect only oscillations around a steady state.

\begin{figure}
    \centering
    \includegraphics{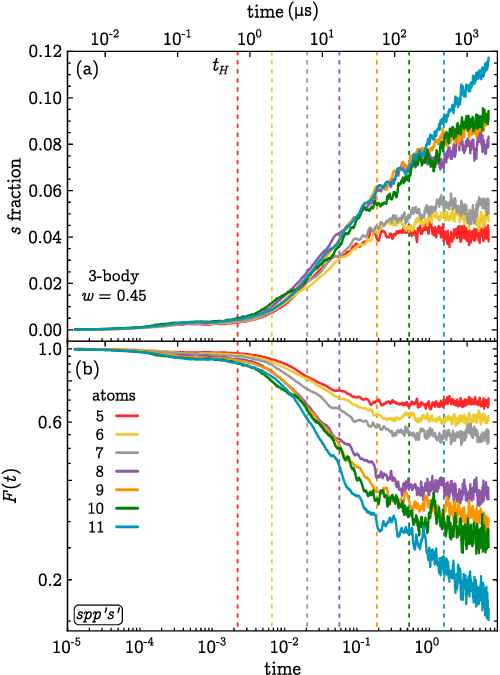}
    \caption{(a) $s$ fraction and (b) fidelity as a function of natural time (lower axis) and real time in $\mu$s (upper axis) for three-body field-tuned interactions in the $spp's'$ model, using the Hamiltonian of Eq.~(\ref{eq:ham3-spps}). Each curve is colored for the number of atoms included in the simulation, as indicated by the legend. The vertical dashed lines show the corresponding Heisenberg time for each case. By the end of the simulated time, none of the $s$ fractions have reached the expected saturation value of 0.241. In (b), the fidelities do not collapse to near zero as they do in the two-body interaction results shown in Fig.~\ref{fig:sps-2-body-conv}(b). This is because the system thermalizes slowly due to a combination of quantum many-body scar states and disorder, as described in Sec.~\ref{sec:slow-therm-sps}. Rather, the asymptotic value of the fidelity is given by the reciprocal of the inverse participation ratio. While these results have not converged, we can still fit the slope of the power-law decay in the maximal case of 11 atoms. This yields $\gamma=0.2$, in agreement with the results shown in Fig.~\ref{fig:gammafit}.}
    \label{fig:three-body-conv-spps}
\end{figure}

\begin{figure}
    \centering
    \includegraphics{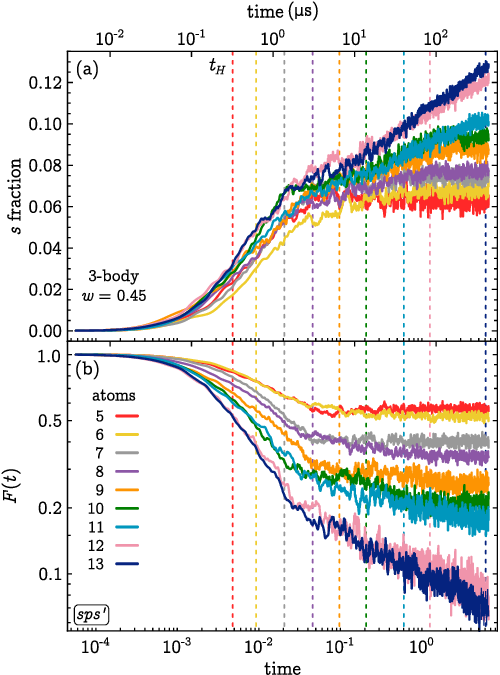}
    \caption{(a) $s$ fraction and (b) fidelity as a function of natural time (lower axis) and real time in $\mu$s (upper axis) for three-body field-tuned interactions in the $sps'$ model, using the Hamiltonian of Eq.~(\ref{eq:ham3-sps}). Each curve is colored for the number of atoms included in the simulation, with their respective Heisenberg times shown by the vertical dashed lines. In (a), it is plausible that the $s$ fractions for the 12 and 13 atom cases have converged. By the end of the simulated time, none of the $s$ fractions have reached the expected saturation value of 0.213. In (b), the fidelities do not collapse to near zero as they do in the two-body interaction results shown in Fig.~\ref{fig:sps-2-body-conv}(b). For 13 and 14 atoms, the fidelities also appear to have converged, in part because the asymptotic values are similar. There is also a sufficient region of power-law decay to attain a robust fit, yielding $\gamma = 0.2$. This result agrees with the $spp's'$ model of Fig.~\ref{fig:three-body-conv-spps}.}
    \label{fig:three-body-conv-sps}
\end{figure}

For the cases with eight atoms and more, $t_H$ is long enough that we can see the system enter the long-time power-law decay phase of its evolution. For thermalizing systems like this, the initial state is spread across the many-body eigenstates, as shown previously in Fig.~\ref{fig:ldos}(a)-(b). If we assume that the initial state is uniformly spread, we can estimate the long-time average of the fidelity to be the reciprocal of the Hilbert space dimension~\cite{tavora_inevitable_2016,tavora_powerlaw_2017}. This estimate works well for the data of Fig.~\ref{fig:sps-2-body-conv}(b). For example, the Hilbert space dimension for eight atoms is 1107, yielding $9\times 10^{-4}$ as an estimate of the asymptotic value of the fidelity. The actual value from the graph is around $7\times 10^{-4}$, as shown by the eight-atom line at late times. Likewise, the Hilbert space dimension for six atoms is 141, yielding $7\times 10^{-3}$ for the asymptotic value of $F(t)$. The actual value from the graph is also about $7\times 10^{-3}$, as shown by the six-atom line at late times.

\begin{figure}
    \centering
    \includegraphics{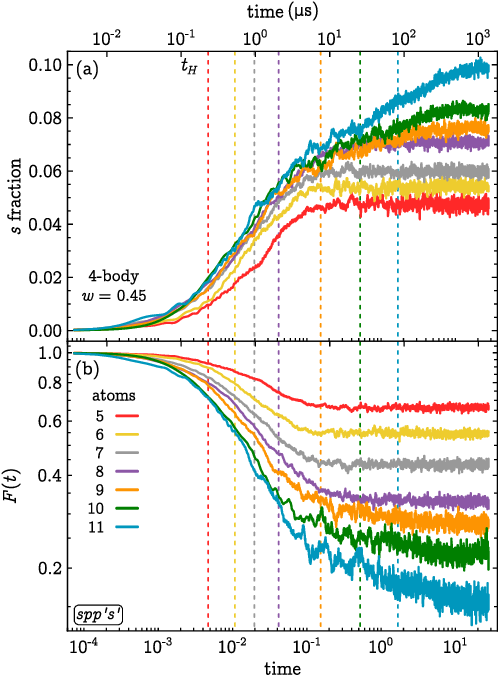}
    \caption{(a) $s$ fraction and (b) fidelity as a function of natural time (lower axis) and real time in $\mu$s (upper axis) for four-body field-tuned interactions in the $spp's'$ model, using the Hamiltonian of Eq.~(\ref{eq:ham4-spps}). Each curve is colored for the number of atoms included in the simulation, as indicated by the legend. The vertical dashed lines show the corresponding Heisenberg time for each case. By the end of the simulated time, none of the $s$ fractions have reached the expected saturation value of 0.193. In (b), the fidelities do not collapse to near zero as they do in the two-body interaction results shown in Fig.~\ref{fig:sps-2-body-conv}(b). Given the values of the Heisenberg times and the fact that all of the fidelities at least begin to plateau, it is difficult to find a region to fit the slope of the power-law decay even for the 11 atom case. Regardless, the slopes are at least consistent with the four-body results presented in Sec.~\ref{sec:slow-therm-sps}, with $\gamma \ll 1$.}
    \label{fig:four-body-conv-spps}
\end{figure}


\begin{figure}
    \centering
    \includegraphics{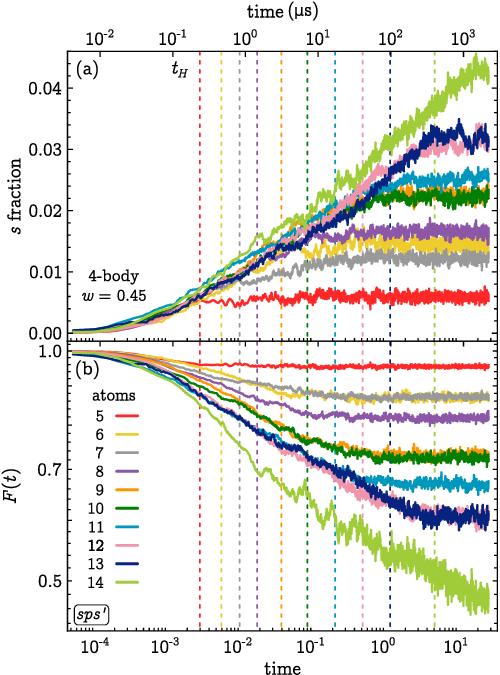}
    \caption{(a) $s$ fraction and (b) fidelity as a function of natural time (lower axis) and real time in $\mu$s (upper axis) for four-body field-tuned interactions in the $sps'$ model, using the Hamiltonian of Eq.~(\ref{eq:ham4-sps}). Each curve is colored for the number of atoms included in the simulation, with the vertical dashed lines representing their respective Heisenberg times. By the end of the simulated time, none of the $s$ fractions have reached the expected saturation value of 0.153. In (b), the fidelities do not collapse to near zero as they do in the two-body interaction results shown in Fig.~\ref{fig:sps-2-body-conv}(b). For both the 13 and 14 atom cases, there is a wide region of power-law decay that allows for a fit prior to or near $t_H$. In both cases, the fit yields $\gamma=0.05$.}
    \label{fig:four-body-conv-sps}
\end{figure}

Considering this fact, one would not expect the fidelity curves in Fig.~\ref{fig:sps-2-body-conv}(b) to overlap at long times as they are each asymptotically approaching a different value due to the different sizes of their Hilbert spaces. If $t_H$ is not sufficiently long, then it is difficult to accurately fit the slope of the long-time power-law decay $F(t)\propto t^{-\gamma}$, as the curve must bend to meet the asymptotic value. For example, for the seven-atom case, there is a region of power law decay before $t_H$. The slope of that decay is artificially smaller than for the 12-atom case, since the seven-atom fidelity plateaus at a higher value. However, if $t_H$ is long enough, the slopes of the curves should converge. This is, in fact, the case for the 11 and 12 atom data. Even though the curves do not overlap, a fit of the slopes yields $\gamma=1.4$ in both cases.

Taken together with the general agreement in the $s$ fraction, the similar power-law decay fits for 11 and 12 atoms in Fig.~\ref{fig:sps-2-body-conv}(b) indicate that our results for the two-body interaction case should be trustworthy. While we cannot claim that the models have converged at 12 atoms, it is plausible that we are near convergence. Furthermore, the Heisenberg times at 11 and 12 atoms are long enough that we can predict thermalization.

Next, we compare the fidelity and $s$ fraction for three-body field-tuned interactions for the $spp's'$ and $sps'$ models, shown in Figs.~\ref{fig:three-body-conv-spps} and~\ref{fig:three-body-conv-sps}, respectively, for $w=0.45$ and $d=9$~$\mu$m. There is no evidence that the $s$ fraction for the $spp's'$ model has converged in Fig.~\ref{fig:three-body-conv-spps}(a). The $s$ fraction at long times for the maximal case of 11 atoms is noticeably different from the $s$ fraction for 10 atoms. However, in the case of the $s$ fraction for the $sps'$ model in Fig.~\ref{fig:three-body-conv-sps}(a), it is plausible that we are nearing convergence. The 12 and 13 atom cases predict nearly the same $s$ fraction, especially for times less than the 12 atom Heisenberg time. Since the number of possible triplets of atoms is much less than the number of possible pairs of atoms at a given total number of atoms, it is reasonable to conclude that more atoms are necessary to achieve convergence for the three-body interactions as compared to the two-body interactions.

In Figs.~\ref{fig:three-body-conv-spps}(b) and~\ref{fig:three-body-conv-sps}(b), the fidelities do not collapse to near zero as in the two-body case of Fig.~\ref{fig:sps-2-body-conv}(b). This is, of course, because these systems thermalize slowly due to a combination of quantum many-body scar states and disorder as discussed in Sec.~\ref{sec:slow-therm-sps}. The initial state is not uniformly spread over the eigenbasis, as shown in the scatter plots of Figs.~\ref{fig:3-4-spps-scatter}(a) and~\ref{fig:3-4-ldos}(a). Rather than using the reciprocal of the size of the Hilbert space to estimate the asymptotic value of the fidelity, one can use the reciprocal of the inverse participation ratio, or IPR, defined in Eq.~(\ref{eq:IPR}). This estimate works very well for the three-body data. For example, in Fig.~\ref{fig:three-body-conv-spps}(b), the estimated asymptotic value of $F(t)$ for the eight atom case is 0.33 and the graph plateaus near 0.35. Similarly, for the eight atom case in Fig.~\ref{fig:three-body-conv-sps}(b) the estimate is 0.37 and the graph plateaus around 0.33.

For the 11 atom case in the $spp's'$ model in Fig.~\ref{fig:three-body-conv-spps}(b), the estimated asymptotic value for the fidelity is 0.11. The fidelity has not yet reached this value, so it is possible that the slope of the long-time power-law decay region has not been too greatly affected. However, not much of the power-law decay occurs before the Heisenberg time and, furthermore, we know from the $s$ fraction analysis that the dynamics have not converged. Nevertheless, we can fit the power-law decay to obtain $\gamma=0.2$. This can be compared to the result shown in Fig.~\ref{fig:gammafit}, which shows the details of the fit for the corresponding three-body $sps'$ result with 13 atoms and also obtains a value of $\gamma=0.2$. This agreement is encouraging.

We can also compare the 12 and 13 atom cases for fidelity in the $sps'$ model shown in Fig.~\ref{fig:three-body-conv-sps}(b). In both cases, there is a large region of power-law decay before $t_H$. The asymptotic values for the fidelity are 0.08 for the 12 atom case and 0.05 for the 13 atom case, which have nearly been reached at the last simulated time. Thus, there is a wide region to obtain a robust fit for the slope. As is visually evident, the two slopes agree and yield the previously calculated value of $\gamma=0.2$. Given this agreement and the possible convergence of the $s$ fraction, our three-body $sps'$ results also seem trustworthy.

Next, we compare the fidelity and $s$ fraction for four-body field-tuned interactions for the $spp's'$ and $sps'$ models, shown in Figs.~\ref{fig:four-body-conv-spps} and~\ref{fig:four-body-conv-sps} for $w=0.45$ and $d=6$~$\mu$m. The $s$ fraction for the $spp's'$ model, shown in Fig.~\ref{fig:four-body-conv-spps}(a), shows only modest evidence of convergence. The $s$ fractions when including nine or more atoms agree fairly well out to about $t=0.1$, near $t_H$ for the nine atom case. It is not surprising that they do not converge at longer times, as the number of possible quadruplets of atoms is greatly restricted even with the maximal case of 11 atoms. In Fig.~\ref{fig:four-body-conv-spps}(b), the $F(t)$ begin to plateau near the asymptotic value estimated from the IPR. Even for 11 atoms, it is not clear that we can obtain a clean fit for the slope of the power-law decay before it plateaus. However, the slopes at least yield $\gamma \ll 1$, in agreement with the results presented Sec.~\ref{sec:fidelity} for the four-body interaction.

The $s$ fraction for the four-body interaction in the $sps'$ model, shown in Fig.~\ref{fig:four-body-conv-sps}(a) shows slightly better, though still modest, evidence of convergence. The $s$ fractions for cases with 12 or more atoms agree fairly well out to about $t=1$, near $t_H$ for the 12 and 13 atom cases. Given the \mbox{two-} and three-body results for the $sps'$ model, it is at least plausible that convergence could be achieved with only a few more atoms. In Fig.~\ref{fig:four-body-conv-sps}(b), the fidelities for all of the cases up to 13 atoms begin to plateau. However, for 13 and 14 atoms there is a significant region of power-law decay before the Heisenberg time and well before the asymptotic plateau of fidelity. Fitting the slopes of these regions, we obtain $\gamma=0.05$ for both cases. This agreement suggests that the $sps'$ model can reliably simulate the long-time evolution of the four-body interactions.

\bibliography{therm-dd}

\end{document}